\documentclass[twocolumn,aps,pre,a4paper,showpacs]{revtex4}
\usepackage{epsfig}

\begin{document}

\title{
Inhomogeneous frozen states in the 
Swift-Hohenberg equation: 
hexagonal patterns vs. localized structures
}
\author{Denis Boyer}
\email{boyer@fisica.unam.mx}
\author{Octavio Mondrag\'on-Palomino}

\affiliation{ Instituto de F\'\i sica, Universidad
Nacional Aut\'onoma de M\'exico, Apartado Postal 20-364, 
01000 M\'exico D.F., M\'exico
}
\date{\today}

\begin{abstract} 
We revisit the Swift-Hohenberg model for two-dimensional hexagonal 
patterns in the bistability region where hexagons coexist with the uniform 
quiescent state. 
We both analyze the law of motion of planar interfaces 
(separating hexagons and uniform regions), and the stability of 
localized structures. Interfaces exhibit properties analogous to that
of interfaces in crystals, such as faceting, grooving and activated
growth or \lq\lq melting".  
In the nonlinear regime, some spatially disordered
heterogeneous configurations do not evolve in time.
Frozen states are essentially composed of extended polygonal 
domains of hexagons with pinned interfaces, that may coexist with
isolated localized structures randomly distributed in the quiescent 
background. Localized structures become metastable at the pinning/depinning 
transition of interfaces.  
In some region of the parameter space, localized structures shrink 
meanwhile interfaces are still pinned. The region where
localized structures have an infinite life-time is relatively limited.

\end{abstract}
%
%

\maketitle

\section{Introduction}

Spatial disorder is a generic feature of pattern formation. 
Whereas the simplest patterns that appear in numbers of 
physical or chemical systems driven out of thermodynamic equilibrium
are periodic structures (such as stripes, squares or hexagons), 
in practice, many steady
patterns observed in spatially extended systems are more complex and 
disordered at large scale \cite{re:rabinovich00,re:cross93,re:manneville90}.
Why disordered configurations spontaneously form (without the help of 
impurities or any quenched disorder) and do not asymptotically evolve 
with time toward more ordered states is still not fully understood.

For instance, systems where various (say, two) patterns of different 
symmetries coexist can easily lead to disordered states.
A basic example
is Rayleigh-B\'enard convection of non-Boussinesq fluids, where hexagonal 
patterns bifurcate subcritically and are stable along with the uniform 
conductive state within a whole interval of Rayleigh numbers. In bistability 
regions, asymptotic configurations may be either homogeneous (containing 
a single phase) or heterogeneous, with both phases present and
distributed in a non-regular way.
Heterogeneous states have been identified experimentally 
in Rayleigh-B\'enard convection \cite{re:boden91,re:assen96}, vibrated
granular materials \cite{re:melo95}, or
Turing patterns and related gas-discharge systems \cite{re:astrov97}.
Spatial disorder has been also proposed as a generic feature of pattern 
formation in ecosystems, as illustrated
by vegetation patches in arid regions \cite{re:lejeune02}.

On the theoretical side, the usual nonlinear model equations 
for periodic patterns often predict metastable disordered states as well. 
Regarding bistable/subcritical patterns, a lot of attention has been devoted 
in the past years to localized structures, that are axisymmetric isolated 
structures of well-defined shape and size
immersed in a uniform quiescent background (of vanishing order 
parameter) \cite{re:coullet87,re:fauve90,
re:gorshkov94,re:tlidi94,re:jensen94,re:sanchez99,re:astrov03}.
Localized structures have been identified as a possible origin of 
disorder in bistable systems, as asymptotic
steady states are in some case composed
of many localized structures randomly distributed in space  
\cite{re:gorshkov94,re:tlidi94,re:lejeune02}. 
The stability of these \lq\lq spatially chaotic" states, as opposed to 
periodic, can be explained by the particular nature of the
effective pair interaction 
between two localized structures \cite{re:coullet87}. 

In the present paper, we consider the Swift-Hohenberg model for hexagonal
patterns in the bistability zone mentioned above and
study spatially more extended objects, namely, islands
with hexagonal order surrounded by the quiescent background.
A localized structure can be seen as a limiting case where such an 
domain is of very small size (of order $\lambda_0$, the wavelength 
of the base pattern). An interesting question to ask is up to what point
localized structures and large islands differ from each other, although
they probably share closely related features.
We therefore analyze in the following the dynamics of a planar interface 
separating an extended domain of hexagons and a uniform phase
in the bistability region. Although interfaces have
been relatively less studied than localized structures in this context
\cite{re:jensen94}, 
we show that they play a central role 
in the large time evolution of two-dimensional extended systems, and in 
the formation of heterogeneous disordered states.

After recalling a few known results on the weakly nonlinear regime
(Section \ref{recall}), we show that in the nonlinear regime
interfaces can either move (with two modes: expanding or shrinking 
domains), or be pinned, depending on the value of the main control parameter 
(Section \ref{ampeq}). The theoretical treatment is based on
an extension of the usual amplitude equation formalism to include
\lq\lq non-adiabatic" effects, according to some ideas presented in
\cite{re:pomeau86,re:bensimon88,
re:malomed90,re:kubstrup96,re:aranson00,re:boyer02}.
Some analogies of this system with interfaces in solid crystals are 
pointed out. The results are further confronted with
numerical resolution of the Swift-Hohenberg equation in
Section \ref{numerical}, where a dynamical diagram for interfaces
is drawn in parameter space.
In Section \ref{disorder}, we discuss the stability of heterogeneous 
frozen states. We show that the stability region of localized
structures only partially overlap the pinning regime of interfaces.
We deduce that the existence of steady disordered 
configurations (and therefore of localized structures of infinite life-time) 
is possible in practice only in that pinning regime.
Disordered frozen configurations obtained from random initial conditions
can be of various types: they can contain localized structures,
or localized structures and extended domains of hexagons (\lq\lq mixed" 
disorder), or extended domains and no localized structures.
Changing the control parameters can drive the system
outside of the pinning zone, keeping localized structures stable:  
these are absorbed by growing domains and therefore become metastable.
Conclusions 
are presented in Section \ref{conclusions}.

\section{Model and known results on interface motion}\label{recall}

We consider a system described by a dimensionless local order parameter 
$\psi(\vec{r},t)$,
representing, for example, the mid-plane vertical velocity in convection 
problems. The spatio-temporal evolution of $\psi$ in non-Boussinesq convection 
can be described by a generalized Swift-Hohenberg equation \cite{re:swift77},
\begin{equation}\label{sh}
\frac{\partial\psi}{\partial t}=\epsilon\psi-\frac{\xi_0^2}{4k_0^2}
(\Delta+k_0^2)^2\psi+g_2\psi^2-\psi^3.
\end{equation}
The parameter $\epsilon\ll 1$ is the main control parameter and 
represents the reduced Rayleigh number $(R-R_c)/R_c$, 
where $R_c$ is the critical Rayleigh number. $k_0\equiv 2\pi/\lambda_0$ 
is the characteristic 
wavenumber of the patterns, and $\xi_0$ the bare coherence length.
In the case of convection, $\xi_0$ is of order $k_0^{-1}$,
its precise value depending on the boundary conditions
at the fluid layer. The coefficient $g_2$ of the quadratic term in 
Eq.(\ref{sh}) models the strength of non-Boussinesq effects; 
$g_2$ will be taken as positive in the following without 
restricting generality ($g_2\ll1$).

Equation (\ref{sh}), or close variants of it, have been studied
in many contexts, for instance, to describe the microphase separation
of asymmetric block-copolymer melts in the weak segregation limit 
\cite{re:bcopol} or
for degenerate optical parametric oscillators \cite{re:tlidi94}. Recently, 
a similar equation was introduced to describe the
evolution of vegetation biomass (the local order parameter 
in that case) in arid climates \cite{re:lejeune02}.

Eq.(\ref{sh}) can be written under a frequently used form:
\begin{equation}\label{sh2}
\frac{\partial\phi}{\partial T}=\Sigma\phi-
(\Delta+k_0^2)^2\phi+G_2\phi^2-\phi^3,
\end{equation}
with 
\begin{eqnarray}\label{rescal}
\phi=(\xi_0/2k_0)\psi,\quad 
T=(\xi_0^2/4k_0^2)t,\quad \nonumber \\
\Sigma=\epsilon/(\xi_0^2/4k_0^2),\quad
G_2=g_2/(\xi_0/2k_0).
\end{eqnarray}
We will take in the following $\epsilon$ and $\xi_0$ 
(rather than $\epsilon$ and $g_2$) as the two main tunable parameters
of Eq.(\ref{sh}), $g_2$ being fixed to an arbitrary small value.
The coherence length $\xi_0$ is thus considered as
independent of $\lambda_0$ {\it a priori}.
From (\ref{rescal}), decreasing $\xi_0$ is equivalent to 
increasing the bifurcation parameters $\Sigma$ and $G_2$.
Keeping $|\epsilon|\ll 1$, the weakly nonlinear regime corresponds
to $\xi_0\sim\lambda_0$, while the nonlinear regime is reached with 
$\xi_0\ll\lambda_0$.

For $\epsilon<0$, the uniform solution $\psi=0$ (conductive state) is 
linearly stable, while it becomes unstable for $\epsilon>0$. 
Standard weakly nonlinear analysis shows that
a subcritical bifurcation to stable hexagonal patterns takes place at
$\epsilon_{\rm m}=-g_2^2/15<0$. Hexagonal patterns remain stable up to 
$\epsilon=16g_2^2/3$, while stripes pattern become stable for 
$\epsilon\ge 4g_2^2/3$ \cite{re:walgraef96}.
In the following, we restrict our study to 
the coexistence zone of the hexagonal and conductive states
($\epsilon_{\rm m}\le\epsilon\le 0$). 

\begin{figure}
\epsfig{figure=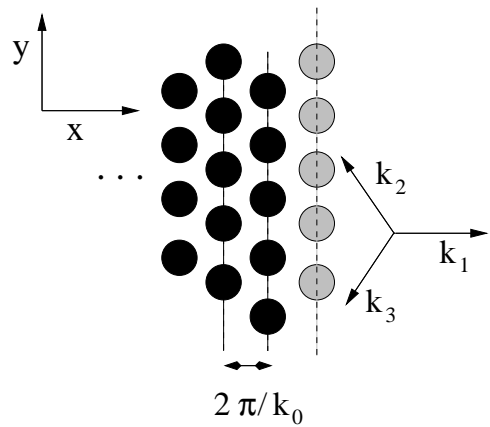,width=2.5in}
\caption{\label{interface} Schematic representation of an interface between 
a hexagonal pattern (left) and a uniform state (right). 
The colors roughly represent the sign of $\psi$. The gray dots correspond to
the next stable position of the interface in the pinning regime 
(see Section \ref{pinning} below).}
\end{figure}

We come back in more details to a problem presented in \cite{re:malomed90}. 
We study a planar interface separating two semi-infinite
domains, one composed of hexagonal patterns and the other in the 
uniform state $\psi=0$. The local order parameter can be approximated at
leading order ($\epsilon^{1/2}$) as
\begin{equation}\label{decomp}
\psi_0=\sum_{n=1}^{3}A_n(x,t)\exp(i\vec{k}_i\cdot\vec{r})+{\rm c.c.}\ ,
\end{equation}
where the amplitudes $A_n$ are slowly varying envelopes depending
on $x$, the coordinate along the axis $\hat{x}$ normal to the 
interface. The solutions of Eq.(\ref{sh}) satisfy the condition
\begin{equation}
\lim_{x\rightarrow-\infty}A_n(x)=A_0>0,
\quad \lim_{x\rightarrow\infty}A_n(x)=0\quad(n=1,2,3). 
\end{equation}
The wave vectors $\vec{k}_n$ are such that
$|\vec{k}_n|=k_0$, and make an angle of $2\pi/3$ with each other. 
We choose $\vec{k}_1$ with an orientation perpendicular to the interface: 
this situation corresponds a \lq\lq faceted" hexagonal pattern
having a raw parallel to the interface, as
schematically displayed in Fig.\ref{interface}.  
Faceted configurations are the more relevant ones, as further 
discussed in Section \ref{numerical}.

When $|\epsilon|\ll 1$, $g_2\ll 1$ and $\xi\sim\lambda_0$,
the equations satisfied by the amplitudes $A_n$ can be derived from 
Eq.(\ref{sh}) by standard multiple scale analysis \cite{re:manneville90}:
\begin{widetext}
\begin{eqnarray}
\frac{\partial A_1}{\partial t}&=&\epsilon A_1+\xi_0^2\partial_x^2A_1
+2g_2\bar{A}_2\bar{A}_3-3(|A_1|^2+2|A_2|^2+2|A_3|^2)\label{amp10}\\
\frac{\partial A_2}{\partial t}&=&\epsilon A_2+\frac{\xi_0^2}{4}\partial_x^2A_2
+2g_2\bar{A}_1\bar{A}_3-3(|A_2|^2+2|A_1|^2+2|A_3|^2)\label{amp20}\\
\frac{\partial A_3}{\partial t}&=&\epsilon A_3+\frac{\xi_0^2}{4}\partial_x^2A_3
+2g_2\bar{A}_1\bar{A}_2-3(|A_3|^2+2|A_1|^2+2|A_2|^2)\label{amp30},
\end{eqnarray}
\end{widetext}
($\bar{A}$ is the complex conjugate of $A$). Note that
$A_3=A_2$ from the planar geometry of the interface. 

We next recall some known results on the velocity $v$ of the
interface. Use the ansatz $A_n=A_n(x-vt)$, multiply 
Eq.(\ref{amp10}) 
(resp. Eqs.(\ref{amp20}) and (\ref{amp30})) by $\partial_t \bar{A}_1$ 
(resp. $\partial_t \bar{A}_2$ and $\partial_t \bar{A}_3$), add 
the equations and integrate their real part over $x$.
The velocity is obtained as \cite{re:malomed90}:
\begin{equation}\label{v0}
v=\frac{A_0^3[45A_0/2-2g_2]}{2D},
\end{equation}
where $A_0$ is the amplitude of the regular hexagonal pattern
\begin{equation}
A_0=\frac{g_2+\sqrt{g_2^2+15\epsilon}}{15},
\end{equation}
and
\begin{equation}\label{D}
D=\int_{-\infty}^{\infty}dx\sum_{n=1}^{3}(\partial_x A_n)^2.
\end{equation}
The numerator of Eq.(\ref{v0}) can be interpreted as a constant driving force
acting on the interface, while $D$ (which depends on the whole interface
profile) plays the role of a friction term.
The system of equations (\ref{amp10})-(\ref{amp30}) has a potential
form: there exists a functional \lq\lq free-energy"
$F\equiv\int d\vec{r}{\cal F}$ such that
$\partial_tA_n=-\delta F/\delta \bar{A}_n$. The driving force in Eq.(\ref{v0})
(plotted below in Fig.\ref{fig:pth} as a function of $\epsilon$)
is proportional to the difference of free-energy density ${\cal F}$ 
between the regular hexagonal state and
the uniform state. Therefore, the sign of $v$ is such that
the domain with the lower free-energy expands at the expense of the 
other. From Eq.(\ref{v0}), $g_2$ being fixed, the velocity changes sign at
$\epsilon=\epsilon_0=-8g_2^2/135=(8/9)\epsilon_{\rm m}$, a value located in the 
coexistence region. The uniform phase spreads if 
$\epsilon_{\rm m}\le\epsilon<\epsilon_0$, whereas 
the hexagonal phase does in the much wider interval 
$\epsilon_0<\epsilon\le0$. 
The interface is stationary only at $\epsilon=\epsilon_0$.
Therefore, except at the marginal value $\epsilon_0$, the amplitude 
equation formalism at leading order predicts that spatially heterogeneous 
distributions of hexagonal and conductive regions should eventually evolve 
toward a single state, either hexagonal or conductive, through interface 
motion.

\section{Non-adiabatic effects}\label{ampeq}

\subsection{Modified amplitude equations} 
 
From Eqs.(\ref{amp10})-(\ref{amp30}), the length scale of variation of
the amplitudes, defining the interface thickness $L$, is roughly of order
\begin{equation}\label{L}
L\sim \xi_0/\sqrt{|\epsilon|}\quad {\rm or}\quad \xi_0/g_2.
\end{equation}
Suppose now that $|\epsilon|$ (or $g_2^2$) is not that small compared to 1, 
and/or $\xi_0$ is significantly smaller than $\lambda_0$:
the assumption $L\gg\lambda_0$ may not be fulfilled in that case. 
The spatial variations of $A_n(x)$ are not necessarily much slower than 
the local oscillations of the patterns, and so-called
\lq\lq non-adiabatic" effects can appear. The effects of 
non-adiabaticity on moving interfaces have been studied 
theoretically for one-dimensional patterns 
\cite{re:bensimon88,re:aranson00}, stripe-stripe interfaces \cite{re:boyer02}, 
square-hexagon interfaces \cite{re:kubstrup96}, and other cases 
\cite{re:malomed90}. Non-adiabatic effects can
be sufficient to pin interfaces and therefore to
stabilize spatially heterogeneous configurations of the order parameter. 
A weakly nonlinear analysis taking into account these effects 
in the present case is reported in Appendix \ref{ap:a}. 
We derive the modified amplitude equations, with (small) non-adiabatic 
corrections that illustrate the coupling between the \lq\lq fast" 
and \lq\lq slow" length scales. Given
the geometry depicted in Fig.\ref{interface}, one obtains
\begin{widetext}
\begin{eqnarray}
\frac{\partial A_1}{\partial t}&=&-\frac{\delta F}{\delta \bar{A}_1}
-i\partial_x\left(g_2A_1^2
-6A_1\bar{A}_2\bar{A}_3\right)\frac{e^{ik_0x}}{k_0}\label{amp12}\\
&+&i\partial_x\left[2g_2(|A_1|^2+|A_2|^2+|A_3|^2)
-6(A_1A_2A_3+\bar{A}_1\bar{A}_2\bar{A}_3)\right]
\frac{e^{-ik_0x}}{k_0}\nonumber\\
\frac{\partial A_2}{\partial t}&=&-\frac{\delta F}{\delta \bar{A}_2}
+i\partial_x\left(2g_2A_2\bar{A}_1-3A_2^2A_3
-3\bar{A}_1^2\bar{A}_3\right)\frac{e^{-ik_0x}}{k_0}\label{amp22}\\
A_3&=&A_2\label{amp32}.
\end{eqnarray}
\end{widetext}
In Eqs.(\ref{amp12})-(\ref{amp32}), the 
$\delta F/\delta\bar{A}_i$'s are the right-hand sides of
Eqs.(\ref{amp10})-(\ref{amp20}). 

The above equations involve directly the space variable $x$ and therefore
break the invariance by translation of standard amplitude equations.
This is a consequence of the coupling between 
the amplitudes and the local phases of the waves. The additional
terms in (\ref{amp12})-(\ref{amp32}) can not be derived from a usual
multiple scale analysis.
Equations (\ref{amp12})-(\ref{amp32}) have been derived in Appendix \ref{ap:a} 
in a lowest mode approximation (valid if $L$ is still $\gg\lambda_0$): 
the non-adiabatic terms oscillate in that case
with the same wavenumber than the base pattern, $k_0$. Other terms 
oscillating as $\exp(2ik_0),\,\exp(3ik_0)...$ should be kept if 
$L\sim\lambda_0$, a situation corresponding to the strongly 
nonlinear regime.

\subsection{Pinned interfaces for $\xi_0$ of the order of $\xi_0^{(p)}$}
\label{pinning}

Fixing the non-Boussinesq parameter $g_2$ to an arbitrary small value,
we determine in this Section the
finite interval of values of the control parameter $\epsilon$ such that 
interfaces do not move, or are pinned. This interval depends on the
coherence length $\xi_0$, and can represent a significant 
part of the coexistence region when $\xi_0$ is small enough.

To derive the modified law of motion of the interface, 
we use the method described in Section \ref{recall}. Denoting 
$x_0$ the position of the interface and using the ansatz 
$A_n=A_n(x-x_0(t))$, one obtains from Eqs.(\ref{amp12})-(\ref{amp32}),
\begin{eqnarray}\label{v}
v=\frac{dx_0}{dt}&=&\frac{F}{D}+\frac{p}{D}\sin(k_0x_0)\label{lawmotion}\\
&\equiv&-\frac{dV}{dx_0}
\end{eqnarray}
with
\begin{equation}\label{force}
F=A_0^3[45A_0/4-g_2],
\end{equation}
and
\begin{widetext}
\begin{eqnarray}\label{p}
p&=&{\rm Max}_{\varphi}\int_{-\infty}^{\infty}dx\cos(k_0x+\varphi)
\left[A_1\partial_x\left(3g_2A_1^2+4g_2A_2^2-18A_1A_2^2\right)\right.
\nonumber\\
&+&\left. 2A_2\partial_x\left(2g_2A_1A_2-3A_2^3-3A_1^2A_2\right)\right],
\end{eqnarray}
\end{widetext}
$D$ being given by (\ref{D}). In Eq.(\ref{v}),
the quantity $p$ is a dimensionless positive number ($p\ll 1$) that 
can be identified with the magnitude
of a periodic pinning potential: the motion of the interface is analogous to
that of a viscous particle sliding down over an 
inclined plane with undulations. From Eq.(\ref{lawmotion}), if 
$p\le |F|$, the interface moves at speed $v\neq0$,
although non-constant in time.
When $p\ge |F|$, the effective potential $V$ seen by the 
interface has many local minima: there exist a set of discrete stable positions 
$x_0^{(p)}$ for which $v=0$. The interface is pinned and hexagonal 
domains do not grow nor shrink.
Two successive steady positions are separated by a distance $\lambda_0$.
This value for the periodicity (result of a single mode approximation, 
see Appendix \ref{ap:a}) has a simple geometrical interpretation: 
the closest steady configuration of the faceted hexagonal domain represented
in Figure \ref{interface} is the one obtained by addition 
(or symmetrically, removal) of an interfacial \lq\lq layer" of dots 
parallel to the interface (the gray dots of Fig.\ref{interface}). 
This situation is analogous to that of a slow solidification (or melting)
process controlled by activation barriers \cite{re:jackson58}.

In expressions (\ref{D}) and (\ref{p}) we have
assumed that the $A_n$'s are real (their imaginary part is supposed to 
be small). 
To evaluate $p$, we solve numerically the first order equations 
(\ref{amp10})-(\ref{amp20}) for the moving interface, 
and substitute in (\ref{p}). Although this solution is only approximate 
(in the pinning regime, we expect some changes in the amplitude profiles 
due to the fact that interfaces are not moving), it captures the correct
physical picture.
A closer comparison with numerical solutions of Eq.(\ref{sh}) is
performed in Section \ref{numerical} below.

\begin{figure}
\epsfig{figure=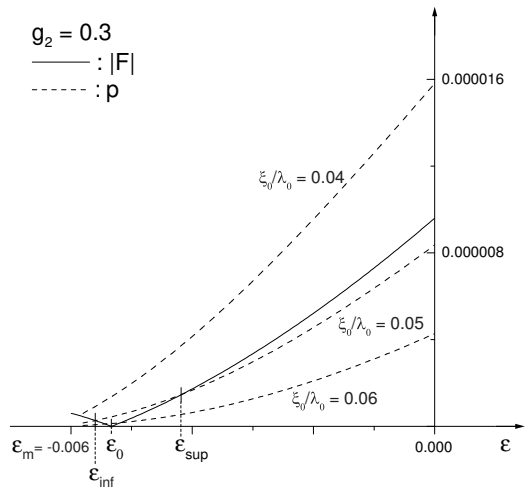,width=2.7in}
\caption{\label{fig:pth} Strength of the driving force $F$ and of 
the pinning potential $p$ (obtained for various values of the coherence 
length $\xi_0$) as a function of $\epsilon$. $g_2=0.3$ here.}
\end{figure}

In Figure \ref{fig:pth}, we plot the driving force $|F|$ and the pinning
potential $p$ given by Eqs.(\ref{force})-(\ref{p})
as a function of $\epsilon$, for a fixed value of $g_2$. We show some
curves of $p$ obtained for different values of the coherence length $\xi_0$
($F$ is independent of $\xi_0$).
Decreasing $\xi_0$ increases non-adiabatic effects and the pinning potential: 
from Eq.(\ref{p}), $p$ roughly behaves as
\begin{equation}\label{estimp}
p\sim A_0^4\exp(-aLk_0),
\end{equation}
with $a$ a numerical constant of order unity and $L$ given by (\ref{L}).

\begin{figure}
\epsfig{figure=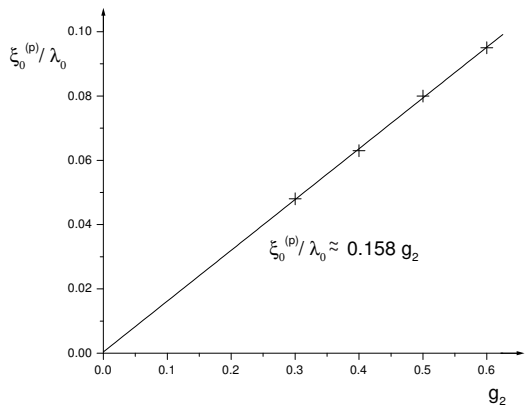,width=2.7in}
\caption{\label{fig:xi0p} Characteristic coherence length $\xi_0^{(p)}$,
defining the onset of a significant pinning regime, as a function of $g_2$, 
as given by weakly nonlinear analysis.}
\end{figure}

At fixed $\xi_0$, interfaces are pinned in a range 
$\epsilon_{\rm inf}(\xi_0)\le\epsilon\le\epsilon_{\rm sup}(\xi_0)$,
while hexagons shrink for $\epsilon_{\rm m}\le\epsilon\le\epsilon_{\rm inf}$
and expand for $\epsilon_{\rm sup}\le\epsilon$, see Fig.\ref{fig:pth}.
For $\xi_0\sim \lambda_0$ (the weakly nonlinear regime), the size of 
the pinning interval
is extremely small compared with $|\epsilon_{\rm m}|$,
and limited to the vicinity of $\epsilon_0$ . 
However, $\epsilon_{\rm sup}-\epsilon_{\rm inf}$ blows up and covers
a significant part of the coexistence region at a cross-over value 
$\xi_0^{(p)}(<\lambda_0)$.
The present calculation predicts that nearly the whole coexistence region 
(at least the interval $[\epsilon_0,0]$) has
pinned interfaces for $\xi_0$ smaller than a critical value
(see Fig.\ref{fig:pth}). 
We define the crossover value $\xi_0^{(p)}$ as given by this critical value. 
In Figure \ref{fig:xi0p}, we plot $\xi_0^{(p)}$ 
as a function of $g_2$. A linear relation is observed:
\begin{equation}\label{xic}
\frac{\xi_0^{(p)}}{\lambda_0}\simeq 0.158\ g_2.
\end{equation}
Since $g_2\ll 1$, $\xi_0^{(p)}\ll\lambda_0$ in Eq.(\ref{xic}). We conclude 
that pinning can be observed in practice only if the scale of variation of 
the amplitude is very short, corresponding to the {\it nonlinear} 
regime of Eq.(\ref{sh}), or equivalently if 
\begin{equation}\label{g2p}
G_2\ge G_2^{(p)}\simeq 2k_0^2
\end{equation}
in Eq.(\ref{sh2}). Unfortunately, in that regime, the equations 
(\ref{amp10})-(\ref{amp30}) used to derive these results are no longer 
valid: higher order terms should be included to improve accuracy.
We therefore do not expect this calculation to give a quantitative 
picture of pinning effects, but rather a qualitative one. 
An example of discrepancy between theoretical and 
numerical results (presented in the following Section) is that the predicted 
pinning in the nearly whole coexistence region at low $\xi_0$
is not observed numerically. Instead, pinning only occurs in the lower 
fraction of the coexistence interval, as further shown in  
Sec. \ref{numerical} below.

\begin{figure}
\vspace{0cm}
\epsfig{figure=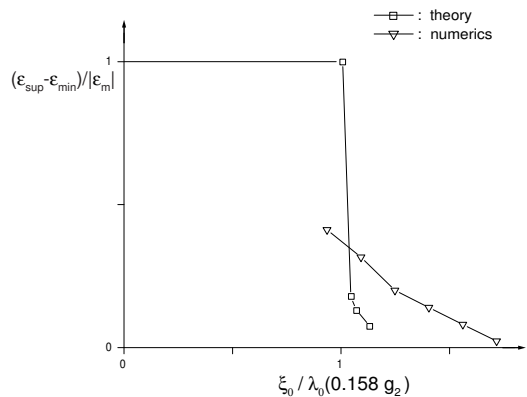,width=2.7in}
\vspace{0cm}
\caption{\label{fig:deps} Relative magnitude of the pinning interval
as a function of $x=\xi_0/(0.158g_2\lambda_0)$, as given by 
weakly nonlinear theory and numerical solution of Eq.(\ref{sh}).}
\end{figure}

Nevertheless, the present weakly nonlinear analysis correctly captures 
the onset of a significant pinning regime for $\xi_0$ below a value 
$\xi_0^{(p)}$, as well as the right order of magnitude for $\xi_0^{(p)}$.
In Figure \ref{fig:deps}, we plot the relative size of the pinning regime
$(\epsilon_{\rm sup}-\epsilon_{\rm inf})/|\epsilon_{\rm m}|$,
determined from theory and numerics,
as a function of $x=\xi_0/(0.158g_2\lambda_0)$.
%
%
Due to the exponential form of Eq.(\ref{estimp}), the theoretical
curve is nearly a step function at $x=1$. The numerical
curve is softer, but with significant variations around $x=1$ as well.

\section{Numerical results}\label{numerical}

\subsection{Pinning of a planar interface}\label{intnum}

We numerically solve Eq.(\ref{sh}) by using a pseudo-spectral method
and a time integration procedure described in \cite{re:cross94}. 
The space is discretized on a square lattice, with $2048$ nodes along 
$\hat{x}$, $128$ nodes along $\hat{y}$, and periodic boundary conditions.
The initial condition is a rectangular hexagonal domain
of length $1028$ and width $128$, 
comprised in-between two quiescent regions of $\psi=0$.
The two interfaces are faceted and parallel to the $\hat{y}$ direction.
The lattice size $\Delta x$ set to unity. The base period 
$\lambda_0= 2\pi/k_0$ of the pattern is $4\sqrt{3}\Delta x$.

\begin{figure}
\vspace{-2cm}
\hspace{-1.75cm}\epsfig{figure=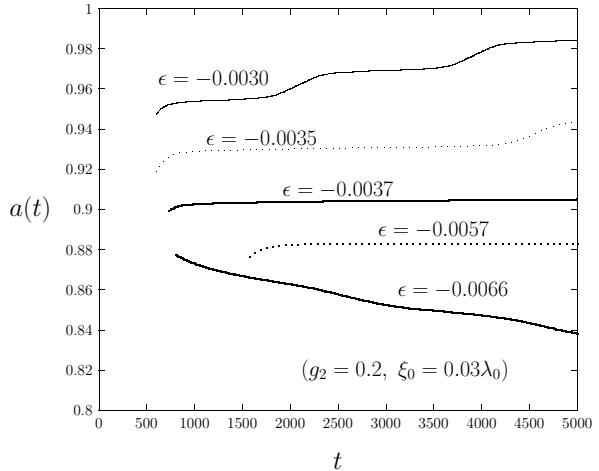,width=4in}
\vspace{-5.5cm}
\caption{\label{position} Position of an interface as a function of
time, obtained by numerical solution of Eq.(\ref{sh}), for various
values of the control parameter $\epsilon$. 
The two upper curves display a net motion over pinning barriers, 
the hexagonal domain expanding into the uniform region 
($\epsilon>\epsilon_{\rm sup}\simeq-0.0037$). The two following 
curves correspond to pinned interfaces 
($\epsilon_{\rm inf}<\epsilon<\epsilon_{\rm sup}$), and the lower one to
shrinking hexagonal domains ($\epsilon_{\rm m}<\epsilon<\epsilon_{\rm inf}$).}
\end{figure}

To determine whether interfaces are pinned or not, we compute the time
evolution of the quantity 
\begin{equation}
a(t)=\int_0^{l_x} \frac{dx}{\lambda_0} \int_0^{l_y} \frac{dy}{l_y}\ \psi^2,
\end{equation}
with $l_x$ and $l_y$ the box dimensions. $a(t)$ is proportional to
the length of the hexagonal domain, and therefore can be used to record
the relative positions of the interfaces. Figure \ref{position} shows
transitions between moving and pinned boundaries as $\epsilon$ is
decreased, for a given $g_2$ and $\xi_0$. As expected, 
hexagonal domains are either expanding, steady or shrinking.
In the moving regimes near a pinning transition, the motion is noticeably 
non-uniform in time, in agreement with the form of the
law of motion (\ref{v}). 

\begin{figure}
\vspace{-2cm}
\hspace{-1.75cm}\epsfig{figure=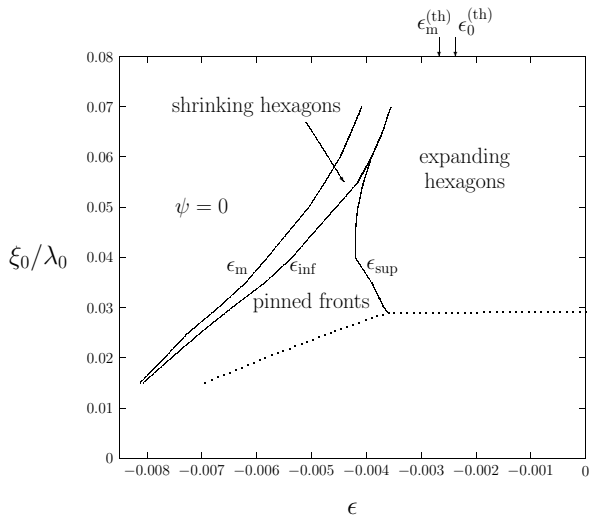,width=4.in}
\vspace{-5.5cm}
\caption{\label{fig:diag.int} Dynamics of hexagons/uniform interfaces
in parameter space $(\epsilon,\xi_0)$, for $g_2=0.2$.
(The shape of the diagram does not depend on the value of $g_2$, that only 
sets the scales of the axis.) The limiting values $\epsilon_{\rm m}^{\rm th}$ 
and $\epsilon_{0}^{\rm th}$ given by weakly nonlinear analysis 
($\xi_0/\lambda_0\sim 1$) are reported 
for comparison. The region below the dotted curve is not of
interest here, see text for details. }
\end{figure}

Figure \ref{fig:diag.int} displays a $(\epsilon,\xi_0)$-diagram representing 
the pinned and moving regimes of hexagon/uniform interfaces. The parameter 
$g_2$ is fixed to 0.2, but the shape of the diagram is independent of this
value, which just sets the scales of both axis
($g_2^2$ for the $x$-axis and $g_2$ for the $y$-axis, 
from Sections \ref{recall} and \ref{ampeq}). For comparison, we also 
report the characteristic values
of $\epsilon$ given by weakly nonlinear analysis ($\xi_0\sim\lambda_0$), 
introduced in Section \ref{recall}, and re-noted 
$\epsilon_{\rm m}^{\rm (th)}$ and 
$\epsilon_0^{\rm (th)}$ here. 
When $\xi_0\ll \lambda_0$, the coexistence region delimited by the
lower value $\epsilon_{\rm m}$ becomes wider as $\xi_0$ decreases:
$\epsilon_{\rm m}$ deviates significantly 
from the value predicted by weakly nonlinear analysis.
Yet, the picture drawn in the previous Section is 
qualitatively reproduced. At any $\xi_0$, one can always identify two 
pinning-depinning transitions:
For $\epsilon_{\rm m}\le\epsilon\le\epsilon_{\rm inf}$, hexagonal domains 
shrink; for $\epsilon_{\rm inf}\le\epsilon
\le\epsilon_{\rm sup}$, interfaces are pinned; for 
$\epsilon_{\sup}\le\epsilon\le 0$, hexagons expand. 
For $\xi_0$ small enough,
the pinning interval $(\epsilon_{\rm sup}-\epsilon_{\rm inf})$ quickly 
grows from almost zero and becomes comparable to $|\epsilon_{\rm m}|$. 
The numerical cross-over value 
found numerically compares well with that given by formula 
(\ref{xic}), as shown in Figure \ref{fig:deps}.
The curve $\epsilon_{\rm m}(\xi_0)$ converges toward
the value $\epsilon_{\rm m}^{\rm (th)}$ as $\xi_0$ increases, while
$\epsilon_{\rm inf}(\xi_0)$ and $\epsilon_{\rm sup}(\xi_0)$ converge
toward $\epsilon_0^{\rm (th)}$. On the other hand, $\epsilon_{\rm m}$
and $\epsilon_{\rm inf}$ tend to merge as 
$\xi_0/\lambda_0\rightarrow 0$, a trend also reproduced by the theoretical
curves of Fig.\ref{fig:pth}.

In the region located below the dotted line of Figure \ref{fig:diag.int}
(low $\xi_0$ and $|\epsilon|$), fronts are moving, with possible instabilities 
toward labyrinthine patterns or non-trivial uniform phases. This region is not 
of interest here and was not investigated.

\subsection{Faceting of oblique interfaces}

One can show from Section \ref{ampeq} that interfaces have to 
be faceted to be pinned. As detailed in the Appendix, if none of the
wavenumbers $\vec{k}_i$ of the base pattern
is directed along the normal to the interface, the nonadiabatic 
terms (\ref{I}) will always integrate to zero along the transverse 
coordinate $y$. In \cite{re:malomed90},  
it was similarly argued that oblique interfaces could not be pinned. 
We show here that oblique interfaces actually modify their shapes
to become pinned. More generally, in the pinning regime,
an hexagonal domain surrounded by a 
conductive state $\psi=0$ evolves toward a polygonal shape
compatible with its three wavenumber directions. This situation 
can be seen as a nonequilibrium analogue of faceting in crystals due 
to strong anisotropy in surface tension \cite{re:herring52}.

\begin{figure}
\epsfig{figure=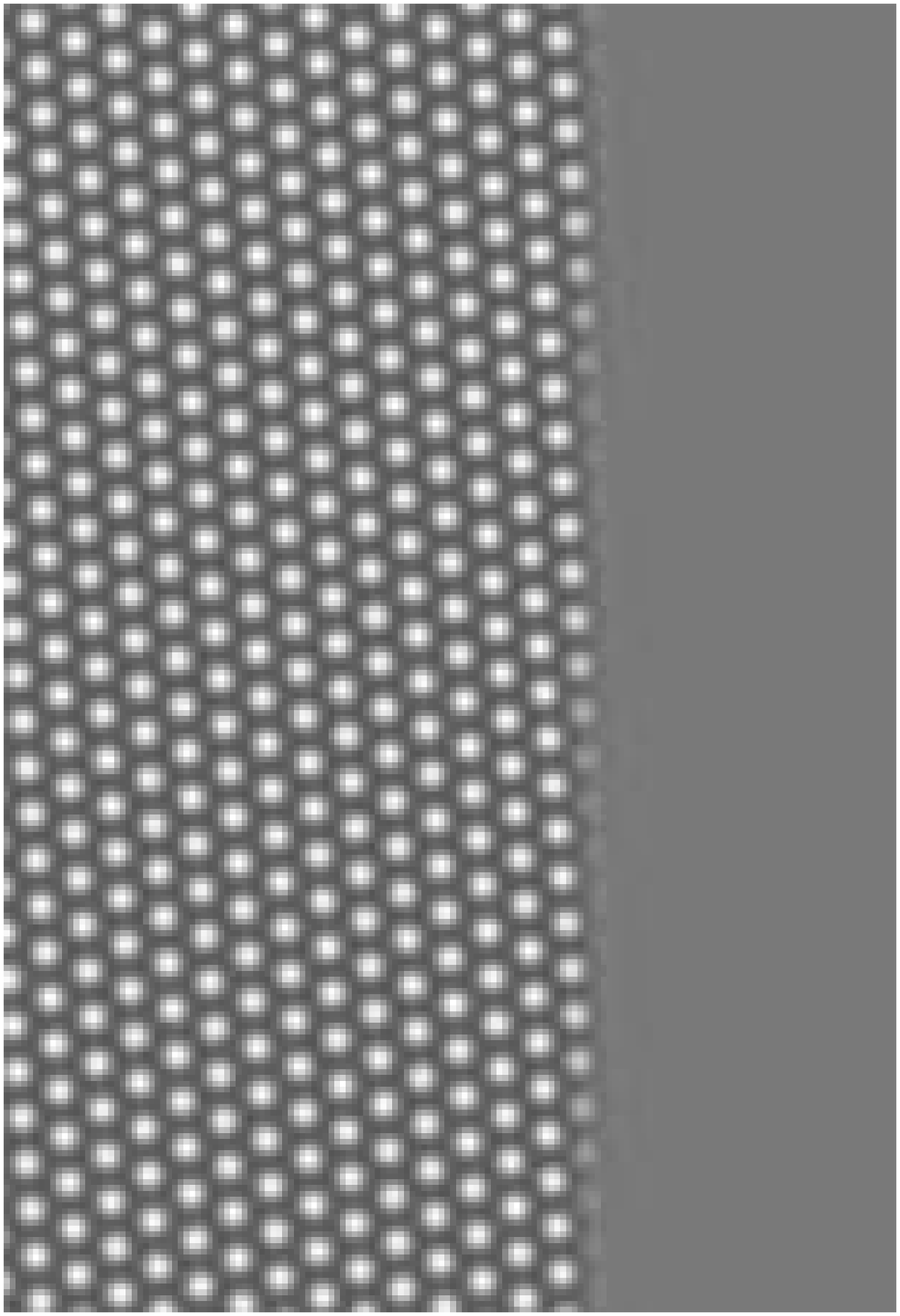,width=1.3in} {\bf a)}
\epsfig{figure=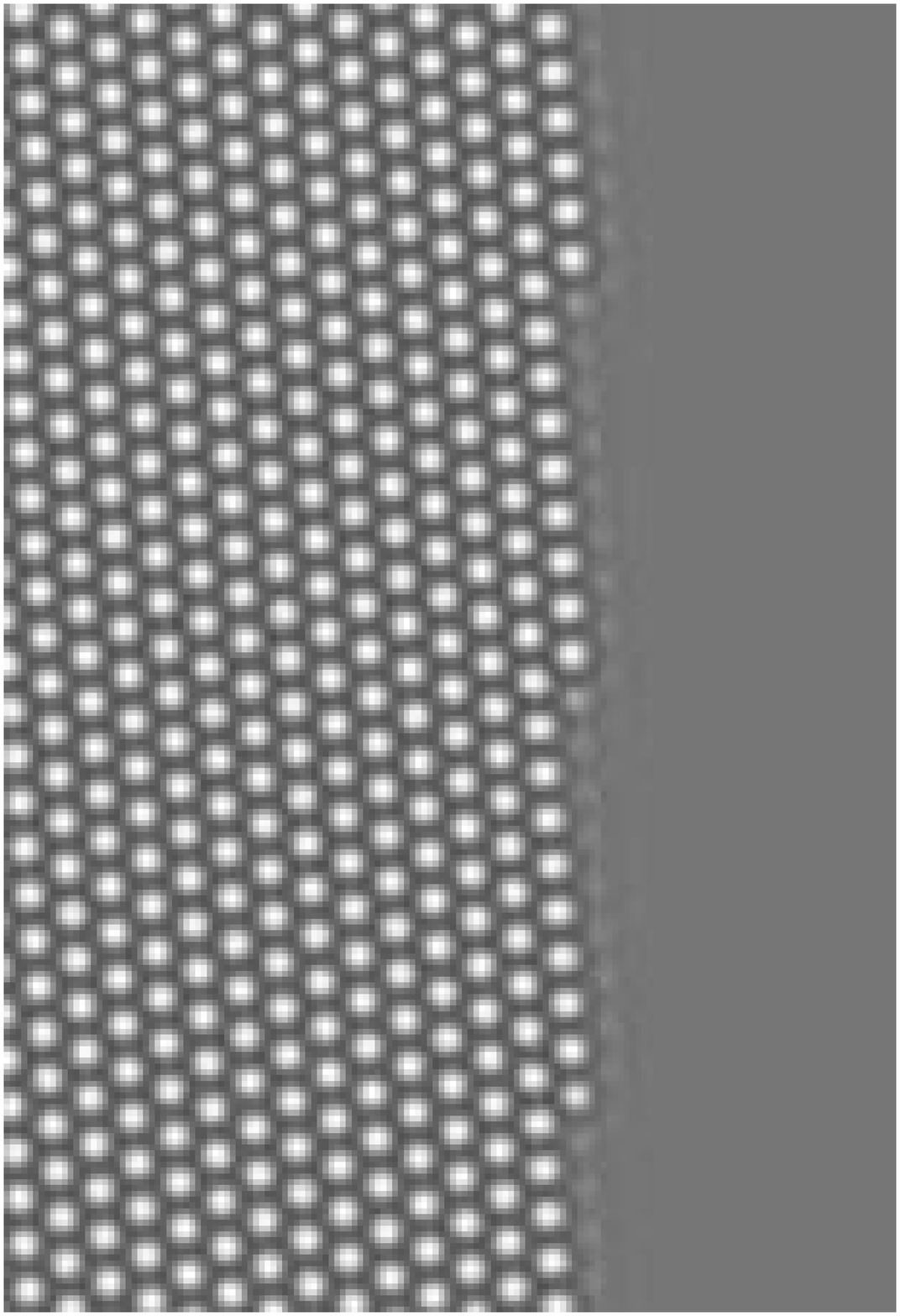,width=1.3in} {\bf b)}
\caption{\label{facet} Faceting of an oblique interface in the pinning
regime ($g_2=0.2,\, \epsilon=-0.0041,\, \xi_0=0.03\lambda_0$). The initial
interface makes an angle $\theta=5.73^{\circ}$ with the normal. a) $t=100$;
b) $t=1900$.}
\end{figure}

As an example,
we have solved numerically Eq.(\ref{sh}) with an initial condition 
composed of a planar interface with a wavenumber $\vec{k}_1$ 
making an angle $\theta=5.73^{\circ}$ with the interface normal 
(Figure \ref{facet}a). The interface evolves toward a wavy, faceted 
shape (Figure \ref{facet}b).

\section{Disordered states: competition between domains of 
hexagonal order and localized structures.}\label{disorder}

From the preceding analysis, we conclude that spatially 
disordered configurations of the order parameter can be stabilized via
pinning effects as soon as the coherence length is of order $\xi_0^{(p)}$
(or $G_2$ in Eq.(\ref{sh2}) of order $G_2^{(p)}$, given by (\ref{g2p})).
These configurations can contain immobile, extended hexagonal domains
with faceted boundaries. In this subsection, we wish to compare the 
previously determined dynamical regimes for planar interfaces with the
stability diagram of localized structures.

Localized structures are axisymmetric (dot-like) structures of well defined 
shape and size frequently encountered in bistability regions.
They have been widely studied and have been recognized to play an important 
role in the production of disordered states 
\cite{re:rabinovich00,re:coullet87,re:gorshkov94}. 
The stability of assemblies of localized structures irregularly distributed 
in space is relatively well understood. The effective interaction 
force between two localized structures changes sign and oscillates 
with their separation distance (a property reminiscent of
Eq.(\ref{v}) for interfaces), therefore allowing many stable discrete 
positions that are unlikely to lead to periodic arrangements of structures 
\cite{re:coullet87}.
We show below that the likelihood of observing localized structures in 
disordered solutions of Eq.(\ref{sh}) in fact largely depends on 
whether planar boundaries are pinned or not. We find that
localized structures can actually coexist with pinned extended domains. 
In other cases, they are metastable, {\it i.e.} susceptible to be swept 
by any expanding domain of hexagonal order.

\begin{figure*}
\vspace{-3cm}
\hspace{-1cm} \epsfig{figure=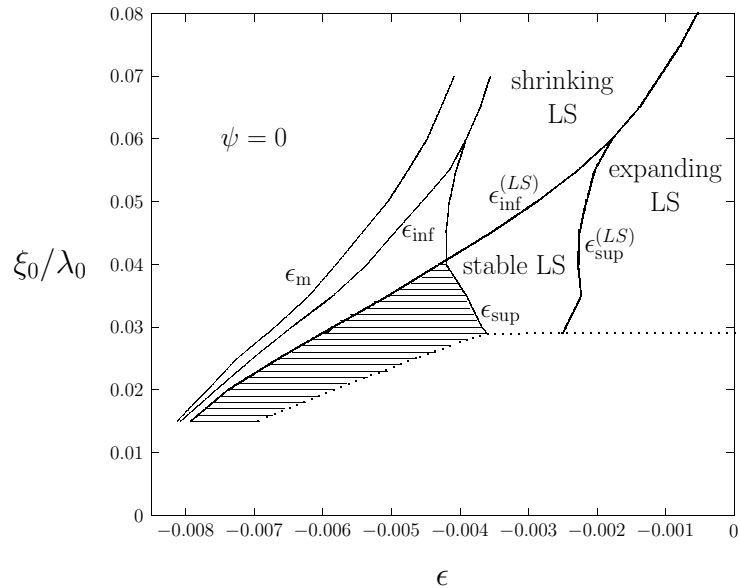,width=5.in}
\vspace{-7.cm}
\caption{\label{fig:diag.int.ls} Stability diagram of localized structures 
(LS), superposed to that of Figure \ref{fig:diag.int} for planar 
fronts ($g_2=0.2$). In the hatched region, fronts are pinned and localized 
structures do not expand nor shrink, and therefore have an infinite life-time.}
\end{figure*}

\begin{figure}
\epsfig{figure=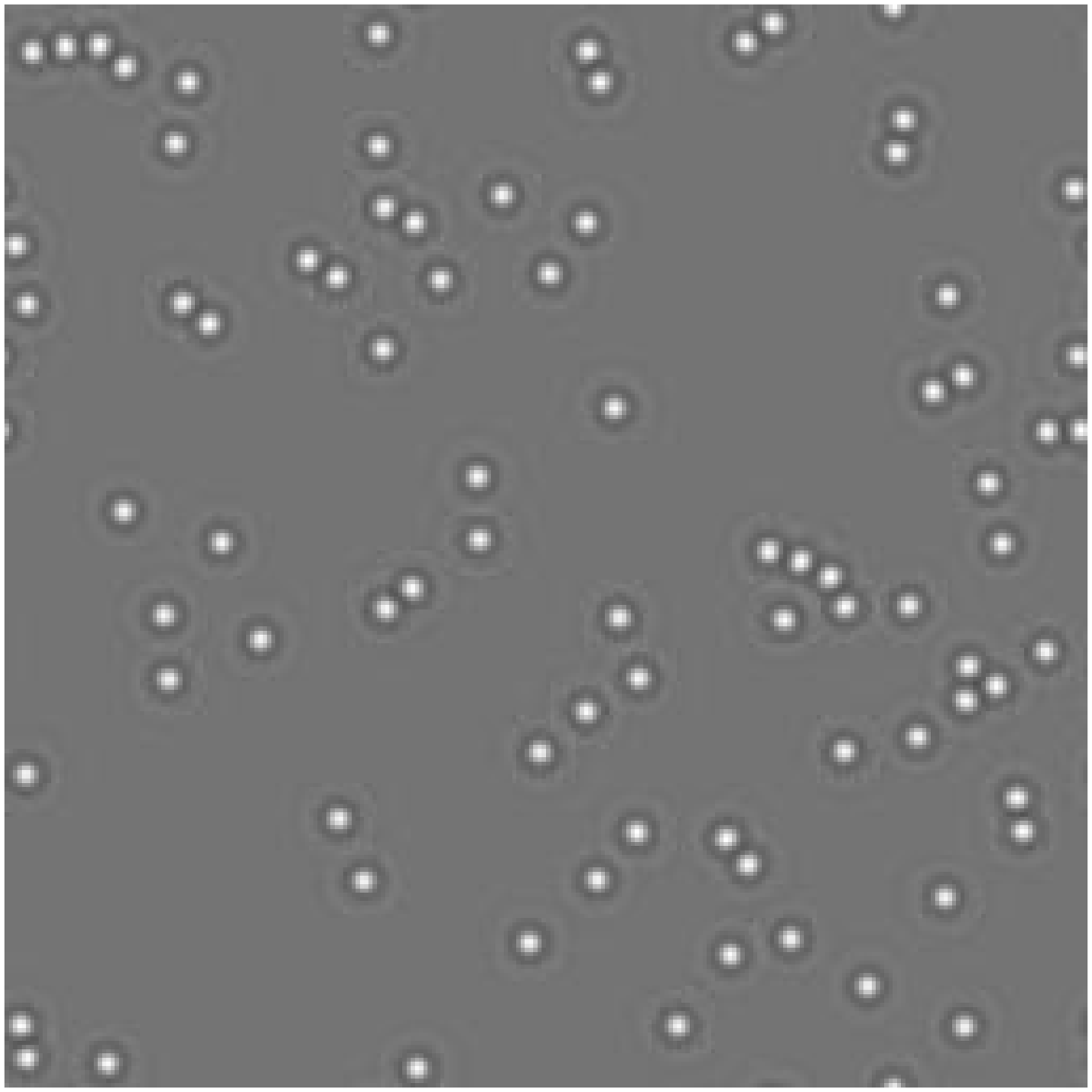,width=1.5in} {\bf a)}
\epsfig{figure=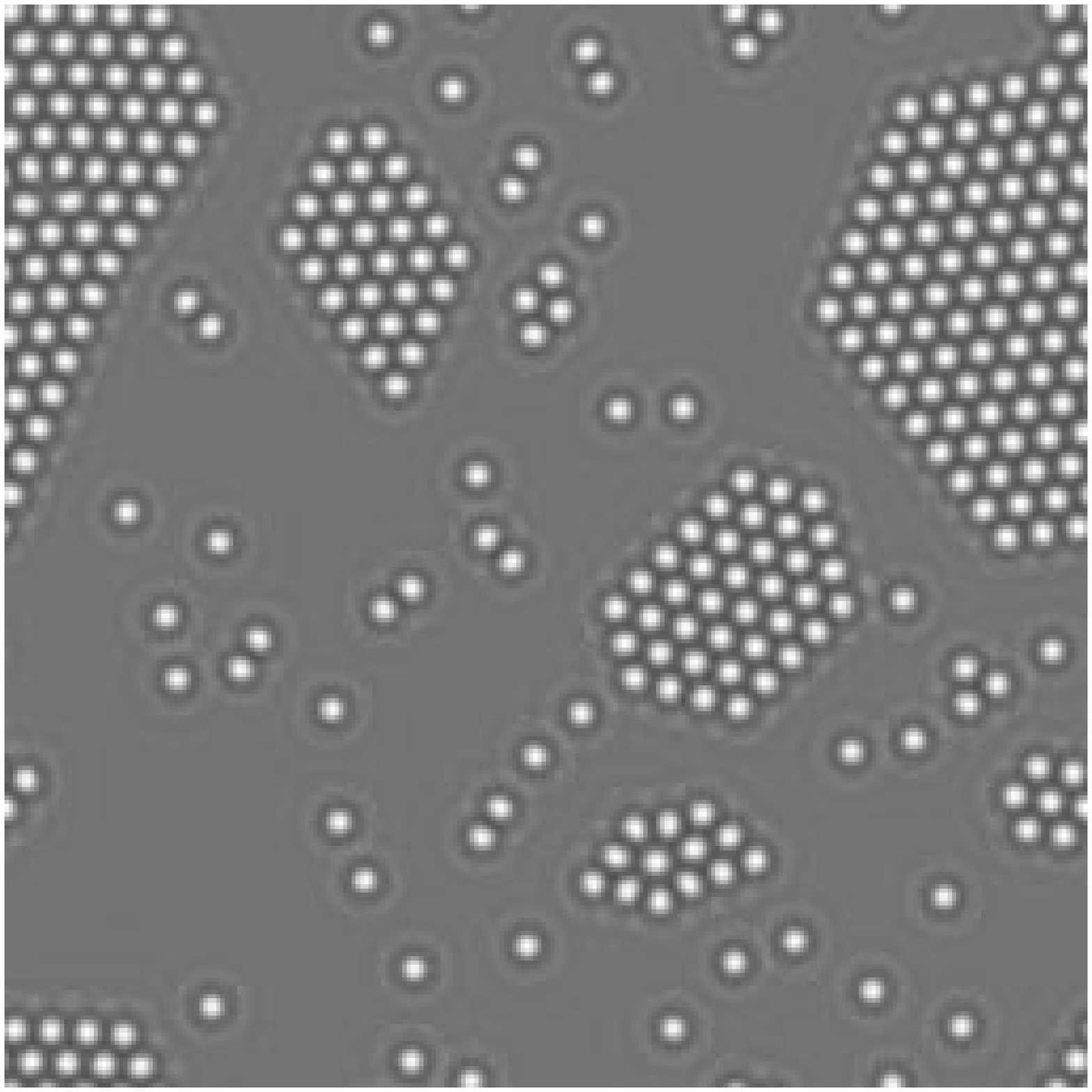,width=1.5in} {\bf b)}

\epsfig{figure=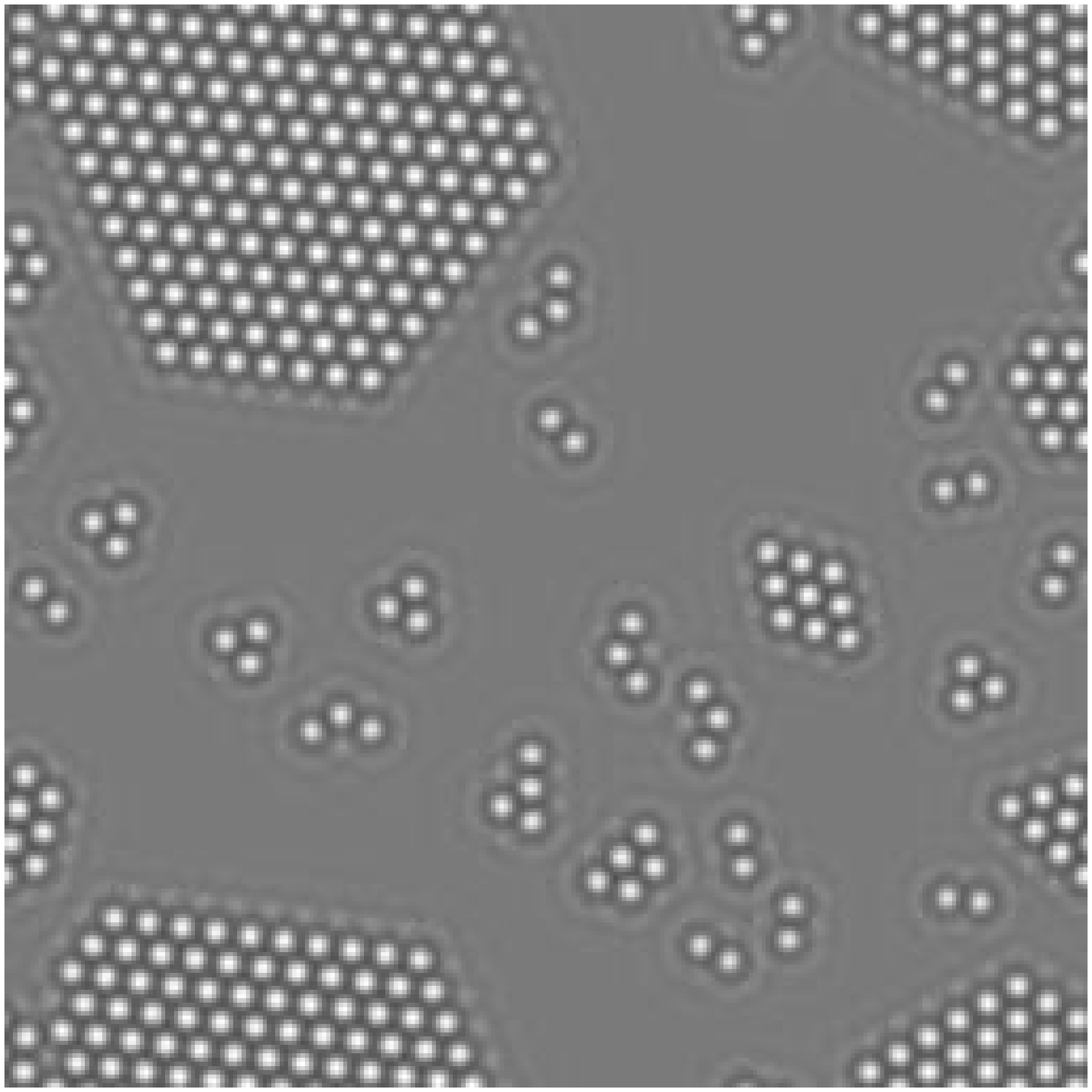,width=1.55in} {\bf c)}
\vspace{0cm}
\caption{\label{frozen} Different kinds of frozen configurations of $\psi$ 
(in gray scale) obtained at large times from random initial conditions 
$(g_2=0.2)$.
{\bf a)} The coherence length is $\xi_0=0.031\lambda_0$ and 
$\epsilon=-0.0055$ is near $\epsilon_{\rm inf}^{(LS)}$ in the hatched 
region of Figure \ref{fig:diag.int.ls}
($t=20000$, $\alpha=13$). {\bf b)} Same $\xi_0$, with
$\epsilon=-0.0040$, closer to  $\epsilon_{\rm sup}$ in the hatched region
of Figure \ref{fig:diag.int.ls}
($t=240000$, $\alpha=8.66$). The pinned hexagonal clusters are 
faceted, and stable localized structures are still observed.
{\bf c)} $\xi_0=0.04\lambda_0$ and 
$\epsilon_{\rm inf}\le\epsilon=-0.0045
\le{\rm Min}[\epsilon_{\rm inf}^{(LS)},\,\epsilon_{\rm sup}]$ : 
interfaces are pinned 
but localized structures are unstable (shrink). 
The smallest stable structures are clusters with two dots.
}
\end{figure}

We have determined the stability diagram for a single localized 
structure. Calculations 
were performed on a square of $128\times128$ grid points over an extended 
period of time (20000 time units), and with a small additive noise term 
in Eq.(\ref{sh}). 
The initial localized structure (a region of size $\sim\lambda_0/2$ 
with $\psi>0$) can either shrink, converge to a stable form, or expand. 
When localized structures expand, they generate hexagonal order, as
described in details in refs. \cite{re:jensen94,re:astrov97}.
Similarly to extended domains, we find that localized structures shrink if 
$\epsilon\le\epsilon_{\rm inf}^{(LS)}$, are stable if 
$\epsilon_{\rm inf}^{(LS)}\le\epsilon\le\epsilon_{\rm sup}^{(LS)}$,
and expand if $\epsilon_{\rm sup}^{(LS)}\le\epsilon$.
The results are reported in the $(\xi_0,\,\epsilon)$-plane in 
Figure \ref{fig:diag.int.ls}, together with
the data of Figure \ref{fig:diag.int} for planar fronts.
Note that the stability zone of localized structures is located 
toward the right of the pinning regime of interfaces.

The stability interval $\epsilon_{\rm sup}^{(LS)}-\epsilon_{\rm inf}^{(LS)}$
and the pinning interval $\epsilon_{\rm sup}-\epsilon_{\rm inf}$ for 
boundaries become
significant in the same range of values of the coherence length $\xi_0$. 
The hatched region of Fig.\ref{fig:diag.int.ls} indicates where
both intervals overlap. In this region of parameters, localized structures
actually have an infinite life-time. Anywhere outside this zone, localized
structure are either unstable, or metastable, {\it i.e.}
susceptible to be swept by a moving front of expanding hexagons. 
Also notice that when $\xi_0/\lambda_0$ is of order 1, 
$\epsilon_{\rm sup}^{(LS)}(\simeq\epsilon_{\rm inf}^{(LS)})$
tends to $0^-$. Therefore, localized structures always shrink in the
weakly nonlinear regime.

To further illustrate these results,
we have solved the Swift-Hohenberg equation (\ref{sh}) with $g_2=0.2$ and
random initial conditions, in various regions of 
Figure \ref{fig:diag.int.ls}.
For convenience, we fix $\xi_0(=0.031\lambda_0)$ in the 
vicinity of $\xi_0^{(p)}$.
The initial condition for $\psi$ is given by a random Gaussian field of 
zero mean and variance $\alpha^2 A_0^2$, with $\alpha$ a dimensionless
tunable parameter of order 10.

Inside the hatched region of Figure \ref{fig:diag.int.ls} 
$(\epsilon_{\rm inf}^{(LS)}\le\epsilon\le\epsilon_{\rm sup})$ and with
$\epsilon$ \emph{relatively closer to $\epsilon_{\rm inf}^{(LS)}$}, the 
asymptotic patterns obtained are mainly composed of aperiodic assemblies of 
localized structures 
of liquid-like character, with no visible hexagonal order.
The example shown in Figure \ref{frozen}a at $\epsilon=0.0055$
(for a system of size $256\times256$)
is similar to the patterns presented in various studies on 
\lq\lq spatial chaos" \cite{re:gorshkov94,re:tlidi94,re:lejeune02}.
Given a random initial condition, at short times, most of the pattern 
quickly relax to the conductive state $\psi=0$, with only a few 
remaining patches that further lead to immobile localized structures.
One can control the final density of localized structures by means of the 
parameter $\alpha$ in the initial condition. 
The configurations become more dilute as $\alpha$ decreases. 

\begin{figure*}
\epsfig{figure=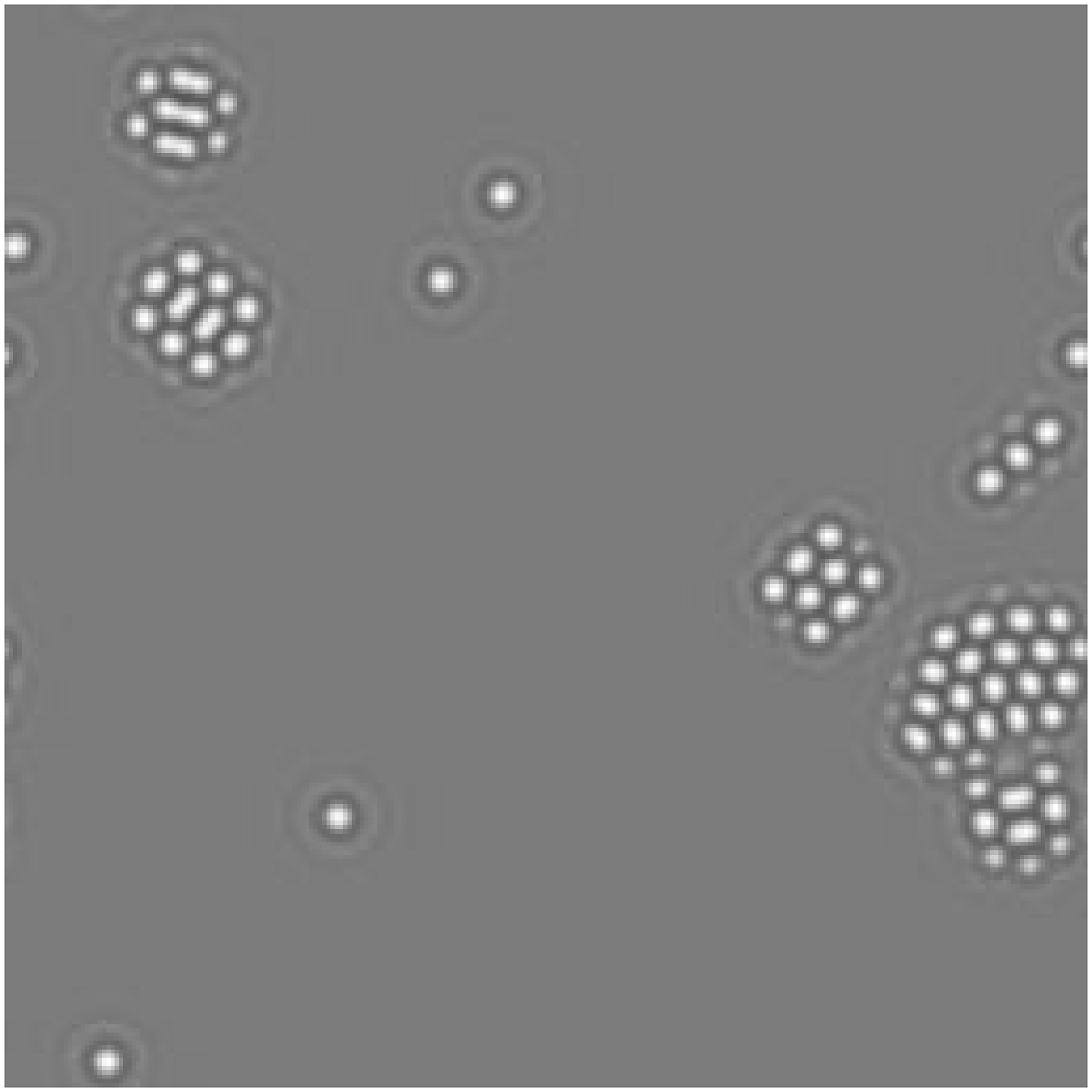,width=1.7in}
\epsfig{figure=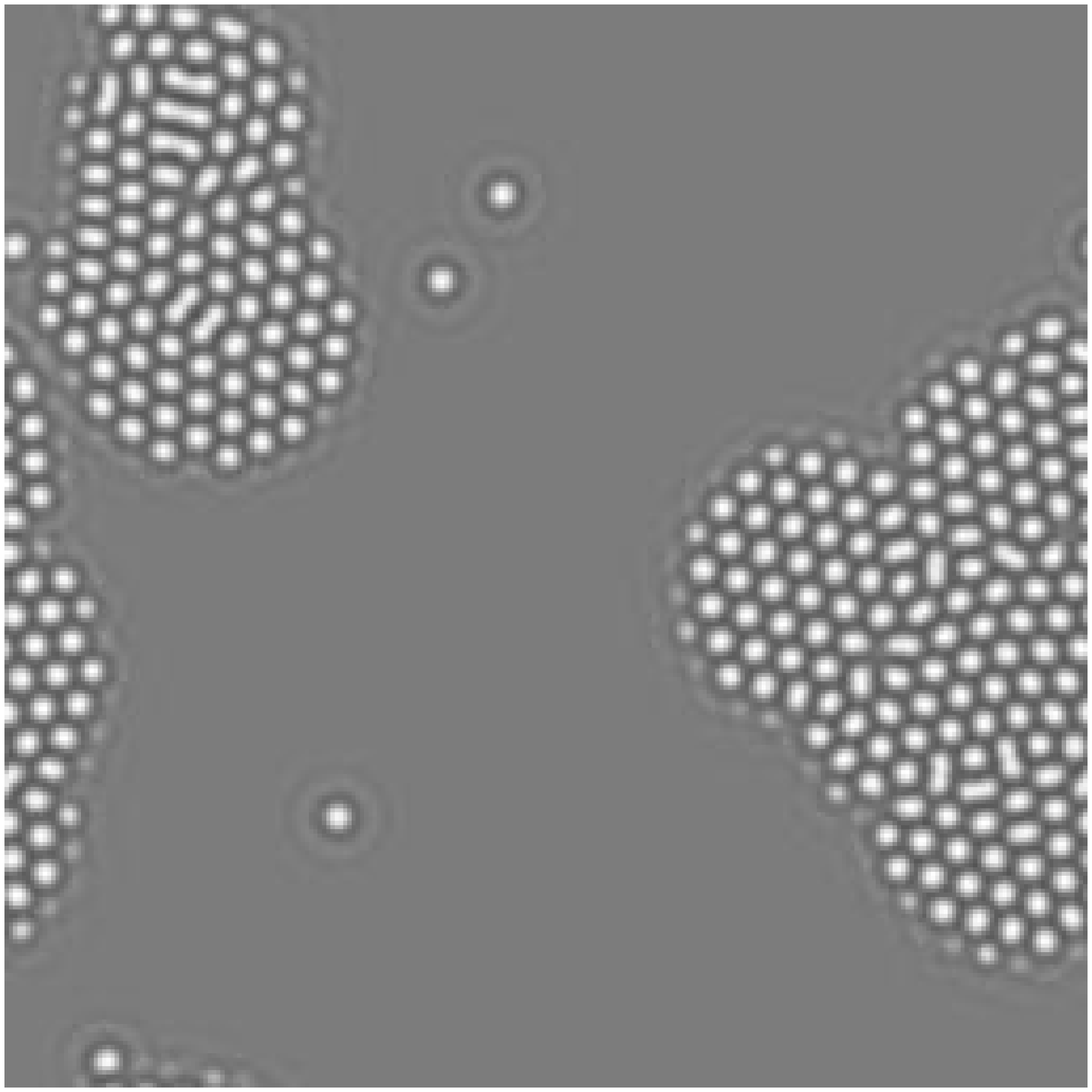,width=1.7in}
\epsfig{figure=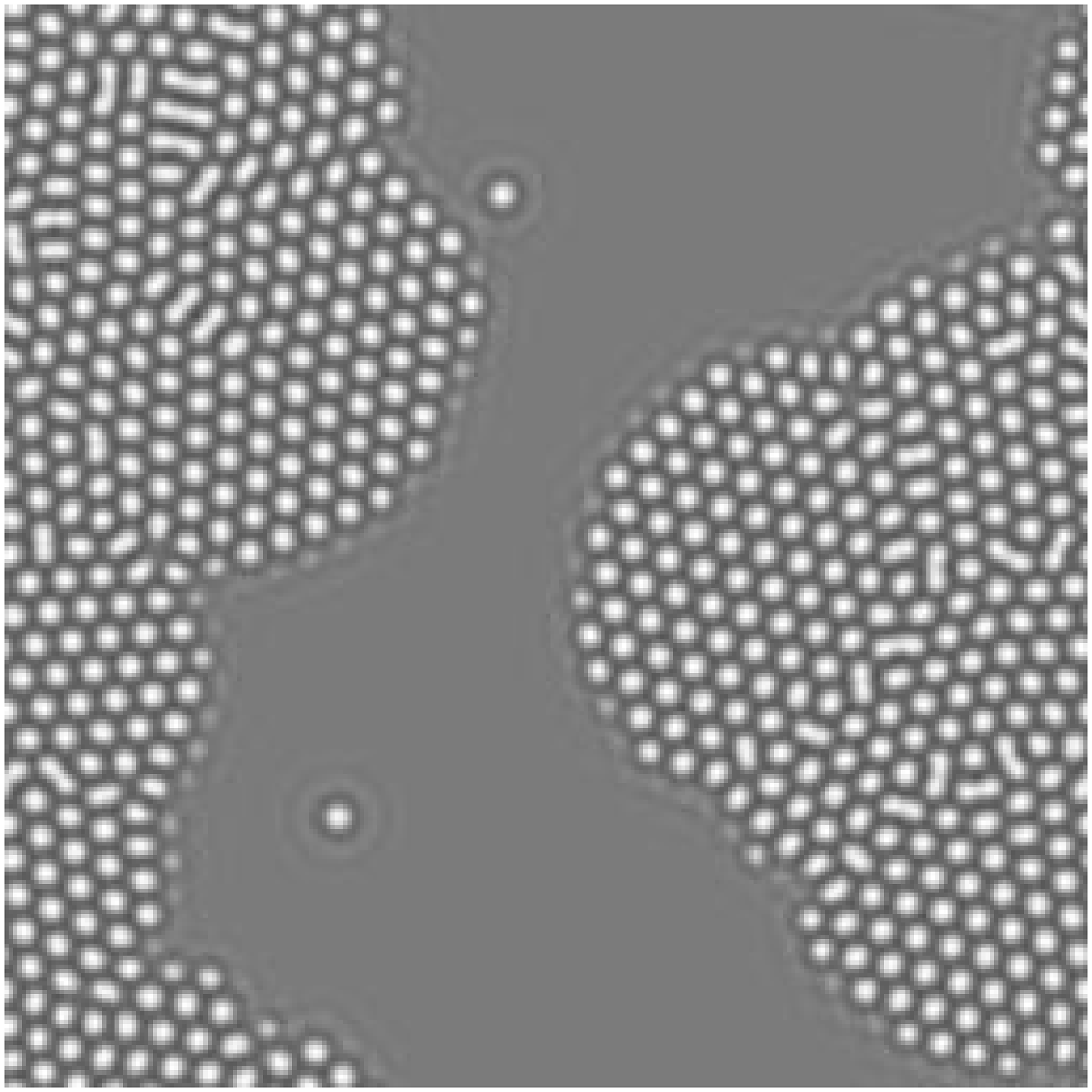,width=1.7in}
\epsfig{figure=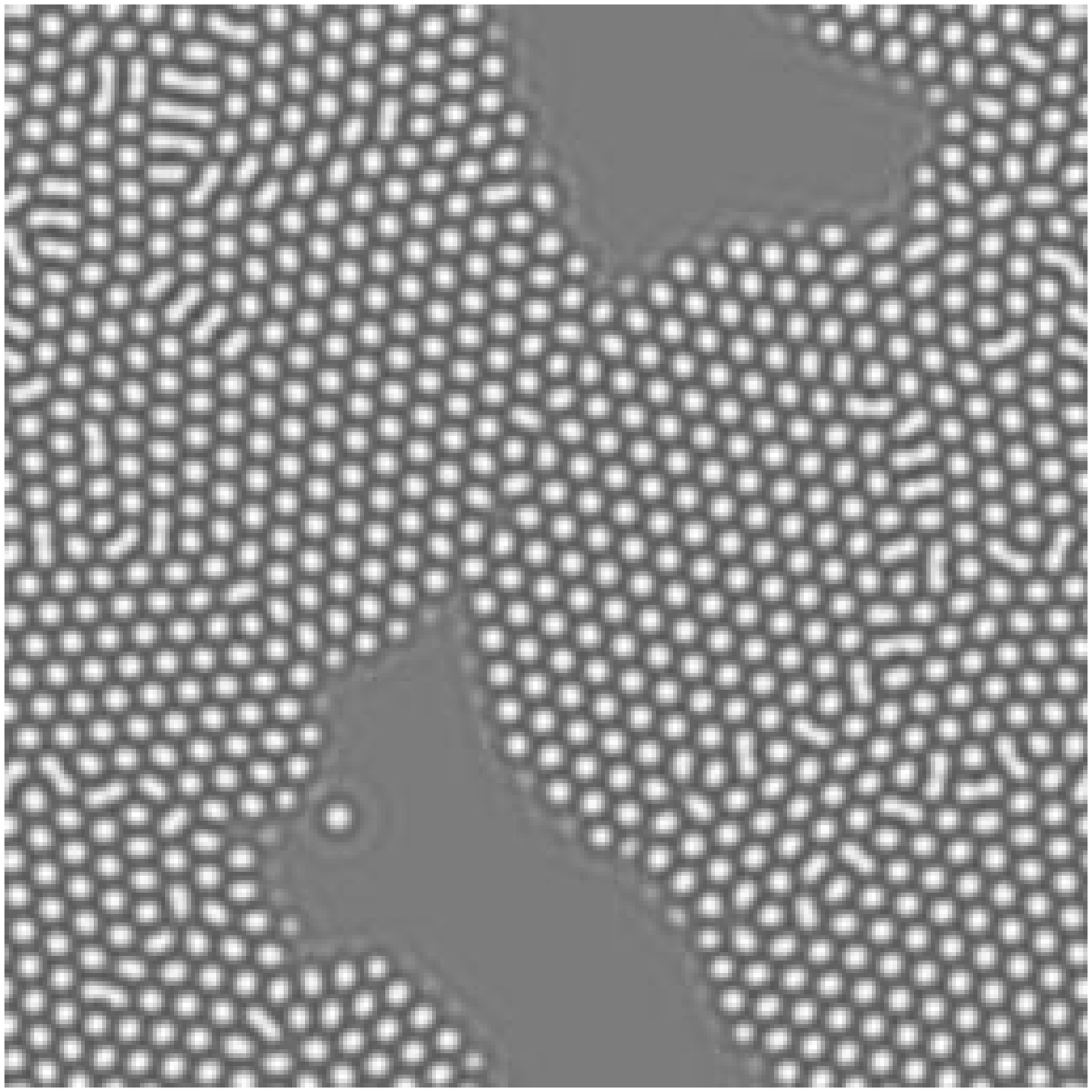,width=1.7in}
\vspace{0cm}
\caption{\label{grow} Pattern evolution for
$\epsilon_{\rm sup}<\epsilon=-0.0026<\epsilon_{\rm sup}^{(LS)}$. (The other 
parameters are $\xi_0=0.031\lambda_0$ and $g_2=0.2$.)
Four successive configurations 
of a same run are shown, respectively at times $t=5000,\, 10000,\, 15000, 
\, 20000$.
Localized structures are stable, but eventually 
absorbed by the growing domains.}
\end{figure*}

\begin{figure*}
\epsfig{figure=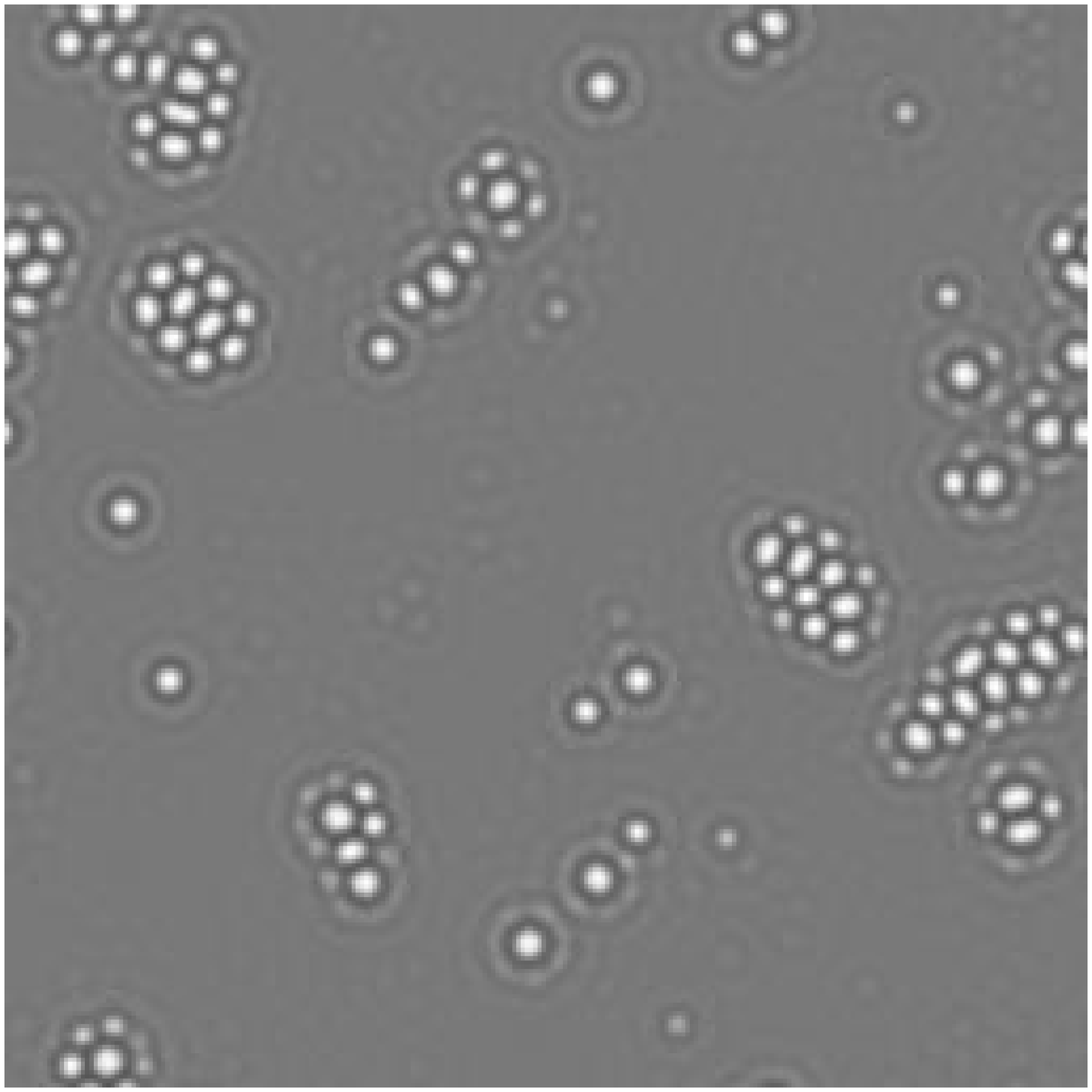,width=2.in}
\epsfig{figure=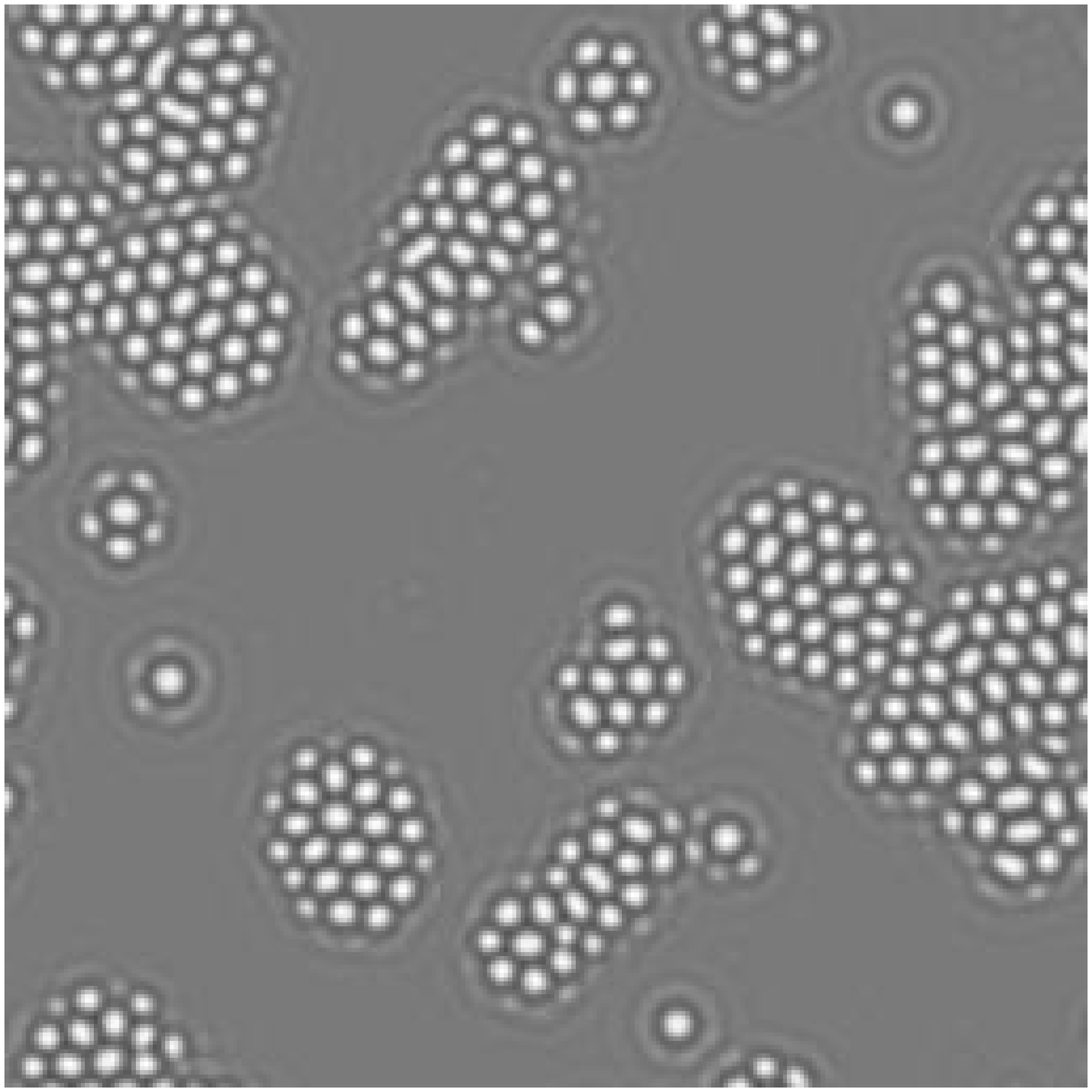,width=2.in}
\epsfig{figure=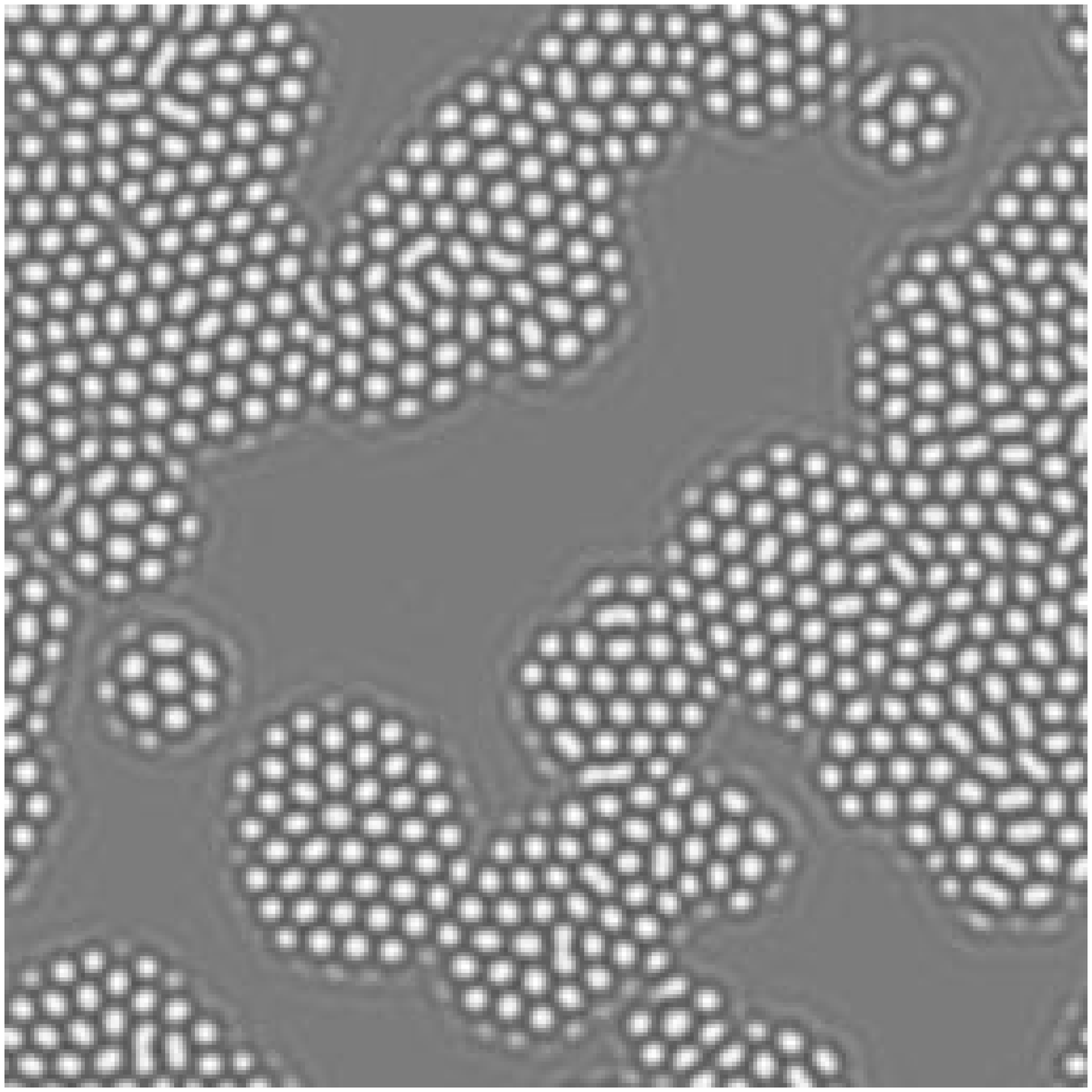,width=2.in}
\vspace{0cm}
\caption{\label{unstable} Same parameters as in Figure \ref{grow}, with
$\epsilon=-0.0010>\epsilon_{\rm sup}^{(LS)}$. Three successive configurations 
at times $t=2000,\, 3000,\, 4000$ are displayed.  
Transient localized structures are now unstable, and nucleate 
fast-growing hexagonal domains.}
\end{figure*}

If $\epsilon$ is increased and set \emph{closer to 
$\epsilon_{\rm sup}$} in $[\epsilon_{\rm inf}^{(LS)},\epsilon_{\rm sup}]$, 
random initial conditions gradually tends to form other kinds of frozen 
disordered states. 
These states are \lq\lq mixed", characterized by 
the presence of both localized structures 
and faceted extended hexagonal domains (Figure \ref{frozen}b).
The interfaces in Fig. \ref{frozen}b are pinned, as expected. 
The enhancement of hexagonal order for quenches closer to
$\epsilon_{\rm sup}$ can be qualitatively understood from 
Section \ref{pinning}: the driving force 
$F$ increases with $\epsilon$ faster than the pinning
potential $p$, even in the pinning zone, see Fig.\ref{fig:pth}. 
The time evolution of a typical configuration is the following:
A few stable localized structures form
at intermediate times. Small clusters, composed of 2,3,4 or more dots, 
are also present. Clusters with 2 dots 
do not grow (nor move). Some clusters with 3 or more dots are unstable: 
some of them, that are initially irregular, expand slowly 
by the creation of successive layers of dots
(they may absorb in the process stable localized structures standing
in their neighborhood).  Growing clusters eventually reach steady states, 
corresponding to faceted polygons with an underlying hexagonal 
structure. These extended hexagonal domains (generalizations of the 7-dots 
structures observed in ref.\cite{re:tlidi94}) 
coexist with the remaining single localized structures. 
It takes a long time (typically $300,000$ times units) for the whole system 
to reach a steady state. The normal to an interface
can be directed along non-crystalline directions, due to, for instance,
merging processes taking place between neighboring clusters.
Domains can be fairly extended by the time they reach a faceted shape,
that further prevent their growth.

The facets ensure stability, although some defects (penta-hepta dislocations) 
can be present in the bulk of a domain. A slight cusp inward is
visible on an interface in the upper left corner of Fig.\ref{frozen}b.
This feature is reminiscent of a grooving effect with sharp angles,
a phenomenon well known in polycrystals with
a strong anisotropy in surface tension \cite{re:wilen95}. 

Further increase in $\epsilon$ beyond $\epsilon_{\rm sup}$ leads, as expected,
to homogeneous states of hexagons, with no conductive regions at large times.
Figures \ref{grow} shows successive configurations obtained at 
intermediate times in the case 
$\epsilon_{\rm sup}\le\epsilon(=-0.0026)\le\epsilon_{\rm sup}^{(LS)}$.
Small hexagonal clusters grow, in a way analogous to a 
stable phase in a metastable one. Although the single localized structures 
present are individually stable, they are absorbed during growth, 
as hexagon/uniform interfaces sweep them in a finite time. Hence,
localized structures can
not be observed at arbitrary large times and are metastable 
in this range. The \lq\lq absolute" stability region of localized structures 
is in practice restricted to the hatched zone of Fig.\ref{fig:diag.int.ls}, 
included in the pinning zone of interfaces. 

When the control parameter exceeds $\epsilon_{\rm sup}^{(LS)}$,
transient short-lived localized structures may still emerge 
from random initial conditions. They further destabilize 
by forming structures of six-fold symmetry and then growing bubbles, 
as illustrated in the sequence of Figure \ref{unstable}.
Localized structures nucleate hexagonal order in this case, as already
observed in the Lengyel-Epstein model of Turing patterns 
\cite{re:jensen94} and experimentally in semiconductor-gas
discharge systems \cite{re:astrov97}.

An other possible case, illustrated by Figure \ref{frozen}c, 
corresponds to pinned interfaces
and unstable (shrinking) localized structures. It happens in the 
roughly triangular region delimited by the curves $\epsilon_{\rm inf}$, 
$\epsilon_{\rm sup}$ and $\epsilon_{\rm inf}^{(LS)}$ in 
Fig.\ref{fig:diag.int.ls}. In Fig.\ref{frozen}c, the smallest structures
observed are clusters with two dots. Finally, for 
$\epsilon\le\epsilon_{\rm inf}$, any initial random condition converges 
toward the conductive state $\psi=0$.

\section{Conclusions}\label{conclusions}

We have studied the dynamics of interfaces in the bistability region 
of conductive and hexagonal states of the Swift-Hohenberg 
equation (\ref{sh}). We have examined the consequences of the
various dynamical regimes of interfaces on the occurrence of spatial
complexity and on the metastability of 
smaller structures like localized structures. 
Steady disordered states are possible in the
nonlinear regime only, and for a choice of
parameters such that planar interfaces remain pinned. 
In the expansion regime 
($\epsilon>\epsilon_{\rm sup}$) localized structure may be stable but
are susceptible to be swept by any traveling interface in a finite time. 

The size of the pinning interval $[\epsilon_{\rm inf},\epsilon_{\rm sup}]$
is significant as soon as the bare 
coherence length $\xi_0$ is of the order of (or smaller than) a 
characteristic value $\xi_0^{(p)}$, approximately given by 
formula (\ref{xic}). 
An extended domain with pinned interfaces can be seen as a nonequilibrium 
analogue of a solid crystal. Domains can contain dislocations
in the bulk, and even vacancies. Defects can induce grooving-like 
effects at the interface (non-planar interfaces with a cusp inward).

On general grounds, decreasing the coherence length 
$\xi_0$ (keeping the other parameters constant) increases nonlinear effects 
and enhances hexagonal order ($\psi(t=\infty)\neq 0$), as illustrated by 
the monotonic curves of $\epsilon_{\rm m}(\xi_0)$ and 
$\epsilon_{\rm inf}(\xi_0)$ in the diagram of Fig.\ref{fig:diag.int}. 
The quantity $\xi_0^{-1}$ can be seen as a measure of
the pattern \lq\lq strength", a \lq\lq weak" pattern
yielding the phase $\psi=0$.
Though, one of the nontrivial features of Fig.\ref{fig:diag.int} is 
the \emph{non-monotonic} behavior of $\epsilon_{\rm sup}$ with $\xi_0$.
There exist a small interval of $\epsilon$'s such that pinned states 
undergo a depinning transition (the system becoming completely 
filled with hexagons) if one slightly \emph{increases} $\xi_0$. 
In this case, a \lq\lq weaker" pattern fills space better.

Heterogeneous frozen
configurations can be of various kinds: they can
include localized structures (\lq\lq liquid"-like disorder), or
both localized structures and extended faceted domains of the condensed
phase (\lq\lq mixed" disorder); or even, faceted domains with small 
clusters and no isolated localized structures. As
the Swift-Hohenberg equation (\ref{sh}) is the minimal model allowing
subcritical patterns and bistability, we expect the results
presented here, specially the conclusions drawn from the
diagrams of Figs.\ref{fig:diag.int} and \ref{fig:diag.int.ls}, 
to be quite general and applicable to a variety of systems. 
Let us mention Turing chemical patterns, nonlinear optical systems 
and related devices \cite{re:astrov97,re:tlidi94}.
(Rayleigh-B\'enard convection and block-copolymer melts are unfortunately
less promising candidates as their $\xi_0$ is relatively large 
\cite{re:manneville90,re:bcopol}.)
%
%
Our results show in particular that crystal-like patterns can be contained 
inside 
polygonal domains compatible with symmetries, and of arbitrary shape. 
Such patterns could be used as lithographic templates 
for technological applications (like micro-electronics) that require 
patterned surfaces over well-defined regions of space.


\begin{acknowledgments}
We thank M. Clerc for fruitful discussions. This work was supported
by the Consejo Nacional de Ciencia y Tecnolog\'\i a (CONACYT, Mexico) 
Grant number 40867.
\end{acknowledgments}

\appendix

\section{Amplitude equations with non-adiabatic effects}
\label{ap:a}

The equations satisfied by the amplitudes $A_n$ can be derived from 
Eq.(\ref{sh}) by standard multiple scale analysis \cite{re:manneville90}.
The spatial variations of $\psi$, see Eq.(\ref{decomp}), involve two kinds 
of length scales: a short scale associated with the periodicity of
modulations, $\lambda_0$, and a large one associated with
the scale of variation of the amplitudes, of order $L$ (the interface
thickness).
When $\epsilon\rightarrow0$ and $g_2\rightarrow0$, $L\gg\lambda_0$,
and it is possible to decompose spatial variations into
\lq\lq fast" variables, $\vec{r}=(x,y)$, 
and a \lq\lq slow" variable $X=\epsilon^{1/2}x$ 
(as well a a slow temporal variable $T=\epsilon t$),
such that $A_n=A_n(X,T)$. The equations for the $A_n's$ are derived from a
solubility condition for $\psi$ at the higher order following $\psi_0$.
The solubility condition for the amplitude $A_n$ reads,
\begin{equation}\label{solvc}
\int_x^{x+\lambda_0}\frac{dx\,'}{\lambda_0}\lim_{l_y\rightarrow\infty}
\int_0^{l_y}\frac{dy\,'}{l_y}[L_n(\psi_{0})
+g_2\psi_{0}^2-\psi_{0}^3]e^{-i\vec{k}_n\cdot \vec{r}^{\,\prime}}=0,
\end{equation}
where the linear operator $L_n=\epsilon-\epsilon\partial_T-
\epsilon\xi_0^2/(4k_0)^2[
2(\hat{k}_n\cdot\hat{x})\partial_x\partial_X]^2$ 
follows from Eq.(\ref{sh}), $\psi_0$ being given by (\ref{decomp})
and $\hat{k}_n=\vec{k}_n/|\vec{k}_n|$. In (\ref{solvc}), we have taken the 
limit $l_y\rightarrow\infty$ for commodity, since the interface profile is 
invariant along $y$. Close to onset, $X$ and $x$ can be considered 
as two independent variables and the only non-vanishing contributions of the
integral (\ref{solvc})
come from the terms oscillating as $\exp(i\vec{k}_n\cdot\vec{r}\,')$ 
in the brackets. Coming back to physical variables, this leads to the 
well-known coupled amplitude equations (\ref{amp10})-(\ref{amp30})

According to relation (\ref{L}), when $\xi_0$ is smaller than $\lambda_0$ 
(or when $\epsilon$ and $g_2$ are not small compared to unity), the 
\lq\lq fast" and \lq\lq slow" variables ($x$ and $X$,
respectively) can not be considered as independent in the solubility
condition (\ref{solvc}) for $A_n$. $\psi_0$ being given by 
the decomposition (\ref{decomp}), if one expands, for instance,
the nonlinear term $g_2\psi_0^2$ in Eq.(\ref{solvc}),
some following integrals appear:
\begin{widetext}
\begin{equation}\label{I}
I=2g_2\int_x^{x+\lambda_0}\frac{dx\,'}{\lambda_0}
\int_y^{y+l_y}\frac{dy\,'}{l_y} A_l(x\,')A_m(x\,')\exp(iK_x x\,')
\exp(iK_y y\,')
\end{equation}
\end{widetext}
with the notation $A_{-l}=\bar{A}_l$ and $-3\le m,l\le3$. 
In Eq.(\ref{I}), $\vec{K}\equiv K_x\hat{x}+K_y\hat{y}
=\vec{k}_l+\vec{k}_m-\vec{k}_n$, with the notation
$\vec{k}_{-l}=-\vec{k}_l$. This time,
let us consider the non-resonant terms only,
{\it i.e.} those with $\vec{K}\neq\vec{0}$.
The $A_n$ profiles of the planar interface depending on $x\,'$ only, 
the contribution $I$ is exactly zero unless $K_y=0$. 
If $K_y=0$ and $K_x\neq0$, then $I$ reduces to
\begin{equation}\label{na1}
I=2g_2\int_x^{x+\lambda_0}\frac{dx\,'}{\lambda_0} A_l(x\,')A_m(x\,')
\exp(iK_x x\,'),
\end{equation}
where $K_x$ is actually a multiple (positive or negative) of $k_0$. 
The usual amplitude formalism assumes that $A_l$ varies slowly,
so that $A_l(x\,')=$cst in the interval $[x,x+\lambda_0]$. In that case,
the non-resonant $I$ given by Eq.(\ref{na1}) always
vanishes. However, as soon as $L/\lambda_0$ is large but finite,
the complex exponential does not exactly integrate to zero if one sits in 
the interface region; $I$, although small, does not vanish.

Anticipating on Section \ref{pinning} (see Eq.\ref{estimp}), 
one can show that a
term like (\ref{na1}) in the solubility condition (\ref{solvc})
gives an additional contribution of order
\begin{equation}\label{estim}
\frac{g_2A_0^3}{D}\exp(-a|K_x|L)\sin(K_x x_{0})
\end{equation}
to the velocity $v$ of the boundary given by Eq.(\ref{v0}). 
(See further Eq.(\ref{lawmotion})-(\ref{p}).) 
In Eq.(\ref{estim}), $a$ is a constant of order unity, $x_0$ the position 
of the interface with respect to an arbitrary origin, and $D$ given 
by (\ref{D}). $|K_x|$ can take the values $k_0$, $2k_0$, $3k_0$... 
In (\ref{estim}), $|K_x|L$ is finite but still assumed to be $\gg 1$:
one can keep among the nonadiabatic terms $I$ those with the
smallest $|K_x|$ ($|K_x|=k_0$), and neglect those involving 
higher wavenumbers.

One now derives the amplitude equations with their
non-adiabatic corrections. One
develops $g_2\psi_0^2-\psi_0^3$ (with $\psi_0$ given Eq.(\ref{decomp})),
and, for each solubility condition (\ref{solvc}) ($n=1,2,3$),  
one has to find out, with the notation introduced above,
which are the products $A_lA_m$ (and $A_jA_lA_m$) such 
that $\vec{K}=\vec{k}_l+\vec{k}_m-\vec{k}_n$ 
(or $\vec{k}_j+\vec{k}_l+\vec{k}_m-\vec{k}_n$) has $K_y=0$ and $K_x=\pm k_0$.
Note that one must eliminate among the eligible products those 
that are resonant in any of the leading order amplitude equations 
(\ref{amp10})-(\ref{amp30}) or their complex
conjugates. These resonant terms, added to similar ones coming from the 
linear operator $L_n$ in (\ref{solvc}), constitute one of the 6
standard amplitude equations: the sum of these terms
cancels out at order $\epsilon^{3/2}$ (the order of the
terms in the brackets of Eq.(\ref{solvc})), and thus would contribute
in non-adiabatic effects at the following order, not considered here.
For example, in deriving the nonadiabatic contributions of the equation for 
$A_2$, one finds an eligible term (\ref{I}) involving a product $A_1A_2$, 
characterized by $|\vec{K}|=|\vec{k}_1|=k_0$. But this term is resonant in 
the equation for $\bar{A}_3$, and hence needs to be discarded. 
The products of amplitudes
appearing in the leading nonadiabatic terms can not be already known
products. 

In order to calculate the integrals
(\ref{na1}), we next assume that the amplitudes
vary slowly in the interval $[x,x+\lambda_0]$ and 
$A_n(x')\simeq A_n(x)+(x'-x)\partial_xA_n(x)$. 
After some tedious but straightforward algebra,
equations (\ref{amp12})-(\ref{amp32}) are obtained.

\bibliographystyle{}

%
%

\end{document}